\pdfoutput=1
\documentclass[a4paper,fleqn,usenatbib]{mnras}

\usepackage[T1]{fontenc}
\usepackage{ae,aecompl}

\usepackage{graphicx}	
\usepackage{amsmath}	
\usepackage{amssymb}	
\usepackage{color}

\definecolor{grey}{rgb}{0.7,0.7,0.7}

\newcommand{\bluetides}{{\sc BlueTides}}
\newcommand{\jwst}{{\em JWST}}

\title[The properties of the first galaxies in \bluetides]{The properties of the first galaxies in the \bluetides\ simulation}

\author[Stephen M. Wilkins et al.]{
Stephen M. Wilkins,$^{1}$\thanks{E-mail: s.wilkins@sussex.ac.uk}
Yu Feng,$^{2,3}$ 
Tiziana Di-Matteo,$^{2}$ 
Rupert Croft,$^{2}$\newauthor 
Christopher C. Lovell,$^{1}$ 
Dacen Waters$^{2}$, 
\\
$^1$\,Astronomy Centre, Department of Physics and Astronomy, University of Sussex, Brighton, BN1 9QH, UK \\
$^2$\,McWilliams Center for Cosmology, Carnegie Mellon University, Pittsburgh PA, 15213, USA \\
$^3$\,Berkeley Center for Cosmological Physics, University of California, Berkeley, Berkeley CA, 94720, USA \\
}

\date{Accepted XXX. Received YYY; in original form ZZZ}

\pubyear{2015}

\begin{document}
\label{firstpage}
\pagerange{\pageref{firstpage}--\pageref{lastpage}}
\maketitle


\begin{abstract}

We employ the very large cosmological hydrodynamical simulation \bluetides\ to investigate the predicted properties of the galaxy population during the epoch of reionisation ($z>8$). \bluetides\ has a resolution and volume ($(400/h\approx 577)^{3}\,{\rm cMpc^3}$) providing a population of galaxies which is well matched to depth and area of current observational surveys targeting the high-redshift Universe. At $z=8$ \bluetides\ includes almost 160,000 galaxies with stellar masses $>10^{8}\,{\rm M_{\odot}}$. The population of galaxies predicted by \bluetides\ closely matches observational constraints on both the galaxy stellar mass function and far-UV ($150\,{\rm nm}$) luminosity function. Galaxies in \bluetides\ are characterised by rapidly increasing star formation histories. Specific star formation rates decrease with redshift though remain largely insensitive to stellar mass. As a result of the enhanced surface density of metals more massive galaxies are predicted to have higher dust attenuation resulting in a significant steepening of the observed far-UV luminosity function at high luminosities. The contribution of active SMBHs to the UV luminosities of galaxies with stellar masses $10^{9-10}\,{\rm M_{\odot}}$ is around $3\%$ on average. Approximately $25\%$ of galaxies with $M_{*}\approx 10^{10}\,{\rm M_{\odot}}$ are predicted to have active SMBH which contribute $>10\%$ of the total UV luminosity.

\end{abstract}


\section{Introduction}

Within the first few hundred million years after the big bang the first stars and galaxies began to form, subsequently bringing an end to the cosmological dark ages. These early galaxies likely produced the ionising photons (either through star formation or from accretion on to super-massive black holes) responsible for the reionisation of hydrogen \citep[e.g.][]{Wilkins2011a, Bouwens2012a, Robertson2013, Robertson2015, Bouwens2015c}.

This critical period of the Universe's history is now observationally accesible thanks largely to the {\em Hubble Space Telescope}. Large samples of star forming galaxies have now been identified at $z\approx 7$ and beyond \citep[e.g.][]{Bouwens2011b, Bouwens2012a, Bouwens2015a, Oesch2010a, Oesch2012b, Bunker2010, Wilkins2010, Wilkins2011a, Finkelstein2010, Finkelstein2012a, Finkelstein2015, Lorenzoni2011, Lorenzoni2013, McLure2011, McLure2013, Ellis2013, Laporte2014, Laporte2015, Laporte2016, Schmidt2014, Atek2015a, Atek2015b}.

Recently, the first small samples have been identified at $z\sim 10$ \citep[e.g.][]{Oesch2012a, Oesch2013b, Oesch2014, Oesch2015b, Oesch2016, Ellis2013, McLeod2015, McLeod2016}, approximately $500\,{\rm Myr}$ after the big bang. By taking advantage of gravitational lensing \citep[e.g.][]{Atek2015a, Atek2015b} and wider-area ground based surveys \citep[e.g.][]{Bowler2014, Bowler2015} galaxies have now been observed at $z\approx 7$ with luminosities spanning a range of four orders of magnitude. The large samples now assembled allow constraints to be placed on the observed UV luminosity function, the star formation rate distribution function \citep[e.g.][]{Smit2012}, and the galaxy stellar mass function \citep[e.g.][]{Grazian2015, Song2015}. 

In the near future, the {\em James Webb Space Telescope (JWST)} will revolutionise the study of the early phase of galaxy formation. \jwst\ will provide sensitive near/mid-IR imaging and spectroscopy with high-spatial resolution. This will potentially allow the discovery of star forming galaxies to $z\sim 15$ and allow the characterisation of the rest-frame UV-optical spectral energy distributions of high-redshift galaxies, providing both accurate redshifts and the robust determination of physical properties (e.g. gas phase metallicities, and star formation rates) through optical emission lines. \jwst\ will be complemented by the Atacama Large Millimetre Array (ALMA) which will provide constraints on rest-frame far-IR emission of galaxies at high-redshift \citep[e.g.][]{Watson2015}. In the longer term the {\em Wide Field Infrared Survey Telescope (WFIRST)} is expected \citep{Waters2016b} to identify large numbers of galaxies to $z\approx 10$ while the upcoming generation of extremely large telescopes will provide much greater spatial and spectral resolution allowing the detailed study of galaxies at very-high redshift.

Through comparisons with predictions from galaxy formation models the observations obtained by {\em Hubble}, and in the future \jwst\ and {\em WFIRST}, will provide the opportunity to test and refine the physics of structure formation, reionisation, and early galaxy formation and evolution. Galaxy formation models can also be used to test and refine observational techniques, for example by assessing how accurately and precisely various physical properties can be recovered  \citep[e.g.][]{Pforr2012, Pforr2013, SH2015} and to influence observational survey strategy; this is particularly important at high-redshift where there are limited existing observations. 

As part of an effort to understand the evolution of galaxies in the $z>8$ Universe and specifically make predictions for both \jwst\ and {\em WFIRST} \citep{Waters2016b} we have carried out a new simulation, \bluetides\ \citep{Feng2015,Feng2016}. \bluetides\ builds upon our previous simulations {\sc MassiveBlack}-I \citep{DiMatteo2012} and {\sc MassiveBlack}-II \citep{Khandai2015, Wilkins2013a} and reaches an unprecedented combination of volume and resolution evolving a $(400/h\approx 577)^{3}\,{\rm cMpc^3}$ cube to $z=8$ with $2\,\times\, 7040^{3}$ particles. \bluetides\ has a resolution comparable to {\sc Illustris} \citep{Vogelsberger2014} and {\sc Eagle} \citep{Schaye2015} but simulates a much larger (approximately $\times190$ larger) volume. 

\bluetides\ also builds upon and complements previous efforts to specifically simulate and model the high-redshift Universe. Recent studies include results from both semi-analytical modelling \citep[e.g.][]{Clay2015, Angel2016, Mutch2016a, Liu2016, Geil2016, Mutch2016b, Liu2017, Cowley2017} and fully hydrodynamical simulations  \citep[e.g.][]{Finlator2006, Finlator2011, Jaacks2012, DiMatteo2012, Johnson2013, Paardekooper2013, Dayal2013, Agarwal2014, Davis2014, Shimizu2014, Khandai2015, Feng2015, Elliott2015, Paardekooper2015, Yajima2015, Feng2016, Jaacks2016, Shimizu2016, Finlator2017, Pawlik2017, Cullen2017}.

In this study we explore the physical and photometric properties of the galaxy population predicted by \bluetides. We begin, in Section \ref{sec:BT}, by describing the \bluetides\ simulation. We then, in Section \ref{sec:physical}, focus on the predicted physical properties of the galaxy population. These include the galaxy stellar mass function (\S\ref{sec:physical.GSMF}), star formation histories (\S\ref{sec:physical.SFH}), and metal enrichment (\S\ref{sec:physical.Z}). In Section \ref{sec:photometric} we present predictions for some of the photometric properties of galaxies, including dust attenuation (\S\ref{sec:photometric.modelling.dust}), average spectral energy distributions (\S\ref{sec:photometric.SED}), and the far-UV luminosity function (\S\ref{sec:photometric.UVLF}). Section \ref{sec:SMBHs} presents predictions for the masses of super-massive black holes and their contribution to the far-UV luminosities of galaxies (\S\ref{sec:SMBHs.UV}). Finally, in Section \ref{sec:c}, we present our conclusions. In Appendix A, we present tables containing the information used to create most of the figures presented in this paper.

\section{The BlueTides Simulation}\label{sec:BT}

The \bluetides\ simulation \citep[\url{http://bluetides-project.org/}, see][for description of the simulation physics]{Feng2015,Feng2016} was carried out using the Smoothed Particle Hydrodynamics code {\sc MP-Gadget} with $2\,\times\, 7040^{3}$ particles using the Blue Waters system at the National Centre for Supercomputing Applications. The simulation evolved a $(400/h\approx 577)^{3}\,{\rm cMpc^3}$ cube to $z=8$ and is the largest (in terms of memory usage) cosmological hydrodynamic simulation carried out. Accompanying the main \bluetides\ simulation was a pathfinder simulation. This evolved a $(50/h)^{3}\,{\rm cMpc^3}$ cube to $z=4$ allowing us to test the simulation predictions against additional observational constraints at $z=4-8$.

Both the main \bluetides\ simulation and the pathfinder were run assuming the {\em Wilkinson Microwave Anisotropy Probe} nine year data release \citep{Hinshaw2013}. Galaxies were selected using a friends-of-friends algorithm at a range of redshifts. At $z=8$ there are approximately 200 million objects identified within the main simulation volume and of these almost 160,000 have stellar masses greater than $10^{8}\,{\rm M_{\odot}}$. At $z=12$ the number of objects identified falls to around 20 million with only around 700 with stellar masses greater than $10^{8}\,{\rm M_{\odot}}$.

\subsection{Comparison with observational constraints}

As demonstrated in \S\ref{sec:physical.GSMF} and \S\ref{sec:photometric.UVLF} the main simulation reproduces current observational constraints on the galaxy stellar mass function and UV luminosity function respectively at $z\approx 8$ and beyond. Using the pathfinder simulation we also find that the good agreement with observed constraints also extends to lower-redshift ($z=4$). However, it is important to note that the pathfinder simulation only simulates a relatively small volume and as such does not provide confirmation of good agreement at high-masses ($\log_{10}(M/M_{\odot})>9.5$).


\section{Physical Properties}\label{sec:physical}

\subsection{Dark Matter - Stellar Mass Connection}\label{sec:physical.DM}

\begin{figure}
\centering
\includegraphics[width=20pc]{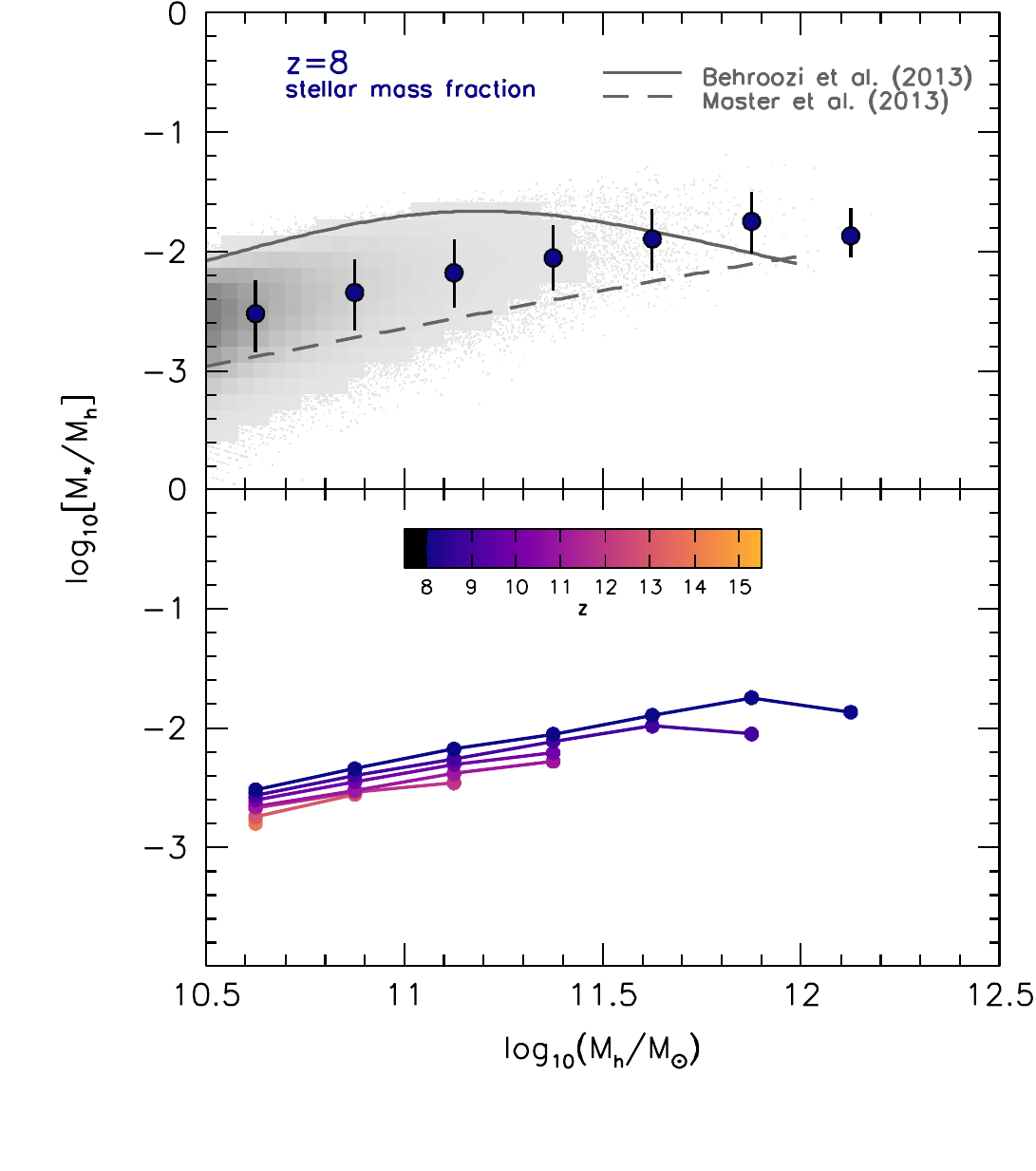}
\caption{The ratio of the stellar to dark matter mass as a function of dark matter mass predicted by \bluetides. The top panel shows the full distribution of sources at $z=8$ with large points denoting the median and the error-bars showing the central $68\%$ range. The lower panel shows only the median values for bins containing $>10$ galaxies at $z\in\{14,13,12,11,10,9,8\}$. Tabulated values of the median ratios are given in Table \ref{tab:DM_stellar}.}
\label{fig:DM_stellar}
\end{figure}

We begin by investigating the link between the dark matter and stellar masses of galaxies predicted by \bluetides. In Fig. \ref{fig:DM_stellar} we show the ratio of the stellar to dark matter masses of galaxies. This ratio increases to higher stellar mass (increasing by approximately 0.5 dex as the dark matter mass is increases by 1 dex) and to lower-redshift. The shape of this relationship broadly matches the extrapolation of the \citet{Moster2013} abundance matching model, however there is a significant difference ($\approx 0.4\,{\rm dex}$) in normalisation. In Fig. \ref{fig:DM_stellar} we also compare our results to the \citet{Behroozi2013} model this time finding a significant difference in both normalisation and shape (at $M_{h}>10^{11}\,{\rm M_{\odot}}$). The exact reason for this is unclear but may reflect that the \citet{Moster2013} and \citet{Behroozi2013} models are calibrated at lower redshift, and thus rely on extrapolation to produce the high-redshift relationship.

\subsection{The Galaxy Stellar Mass Function}\label{sec:physical.GSMF}

\begin{figure}
\centering
\includegraphics[width=20pc]{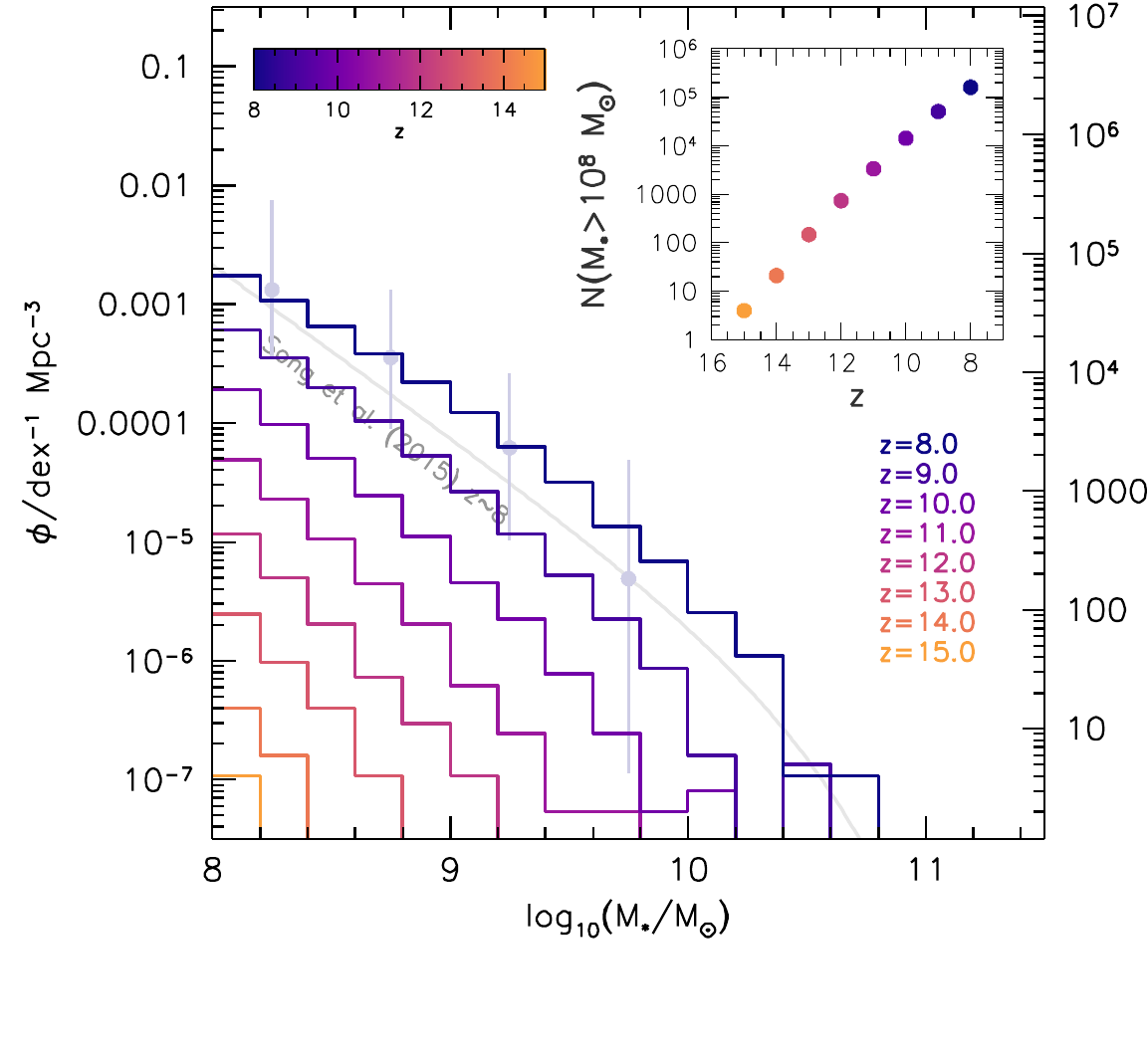}
\caption{The galaxy stellar mass function predicted by \bluetides\ at $z\in\{8,9,10,11,12,13,14,15\}$. The right-hand axis shows the total number of galaxies in \bluetides\ in each $\Delta\log_{10}M=0.2$ mass bin. The points and grey line show observational constraints from \citet{Song2015} at $z\approx 8$ corrected to assume a \citet{Chabrier2003} initial mass function. The inset panel shows the number of objects with $\log_{10}(M_*/{\rm M_{\odot}})>8$ in the simulation volume as a function of redshift $z=15\to 8$. Tabulated quantities from \bluetides\ are given in Table \ref{tab:GSMF}.}
\label{fig:GSMF}
\end{figure}

The galaxy stellar mass function (GSMF) predicted by \bluetides\ is shown in Fig. \ref{fig:GSMF}. At $z=8$ \bluetides\ simulated a sufficiently large volume to robustly model the GSMF to stellar masses of $>10^{10}\,{\rm M_{\odot}}$. From $z=15\to 8$ the number of $>10^{8}\,{\rm M_{\odot}}$ galaxies within the simulation increases from a handful at $z=15$ to almost 120,000 by $z=8$ demonstrating the rapid assembly of the galaxy population during this epoch. Over the period the shape of the GSMF also evolved, with the number density of massive galaxies increasing faster. For example, from $z=10\to 8$ the number density of galaxies with $M_*\approx 10^{9.5}\,{\rm M_{\odot}}$ increased a factor of $\approx 4\times$ faster than those with $M_*\approx 10^{8}\,{\rm M_{\odot}}$.

It is now possible, by combining deep {\em Hubble} observations with {\em Spitzer}/IRAC photometry, to probe the rest-frame UV-optical spectral energy distributions of galaxies at very-high redshift, and thus measure robust stellar masses and thus the galaxy stellar mass function. 

While several studies have constrained the GSMF at very-high redshift \citep{Gonzalez2011, Duncan2014,Grazian2015,Song2015} only \citet{Song2015} have extended observational measurements of the GSMF to $z\approx 8$ overlapping with \bluetides. The \citet{Song2015} results are shown Fig. \ref{fig:GSMF} and closely match the \bluetides\ predictions over much of the simulated and observed mass range. The possible exception to this otherwise excellent agreement is at high masses $M_*>10^{10}\,{\rm M_{\odot}}$ where \bluetides\ appears to predict more galaxies than are currently observed (although the observational uncertainties are very large). While this may reflect modelling issues it is also likely there exist observational biases at these large masses. The most-massive systems are predicted to be heavily obscured, even at $z\approx 8$, and may fall out of UV selected samples.

It is also important to note that there are large differences between the observed GSMFs presented by different studies at very-high redshift. For example, despite using a largely overlapping set of observations \citet{Song2015} find number densities (at $M_{*}>10^{9}\,{\rm M_{\odot}}$) almost an order of magnitude lower than \citet{Duncan2014} - for a discussion of the many issues regarding observational estimates of the GSMF see \citet{Grazian2015} and \citet{Song2015}. Observational estimates of the GSMF are sensitive to the choice of initial mass function (IMF). Assuming a \citet{Salpeter1955} IMF for example would lead to observational mass estimates systematically increasing by approximately $0.17\,{\rm dex}$.


\subsection{The Star Formation Rate Distribution Function}\label{sec:physical.SFRDF}

\begin{figure}
\centering
\includegraphics[width=20pc]{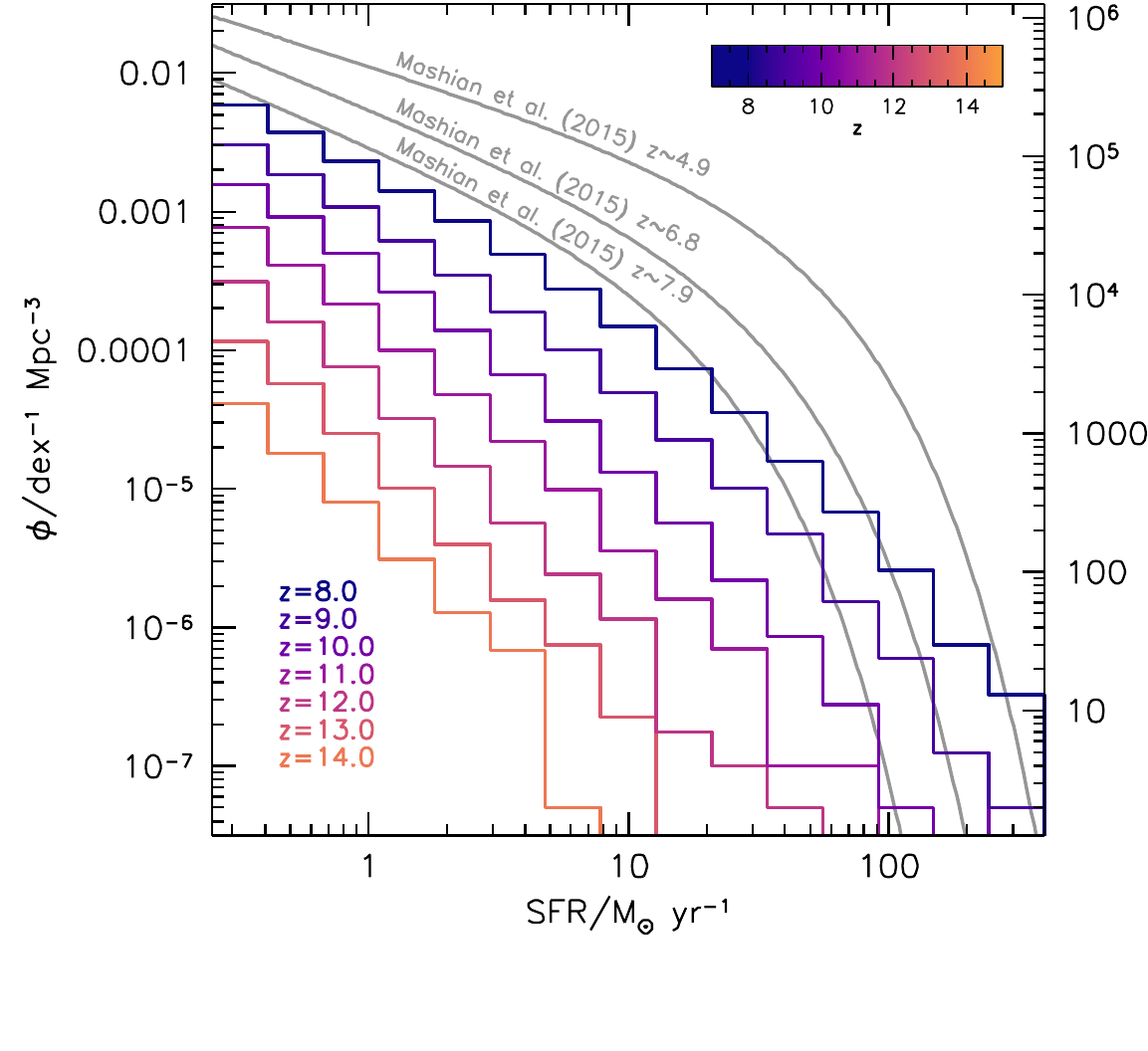}
\caption{The star formation rate distribution function predicted by \bluetides. The right-hand axis shows the total number of galaxies in \bluetides\ in each $\Delta\log_{10}SFR=0.2$ bin. Solid lines show the dust-corrected (intrinsic) star formation rate distribution functions measured by \citet{Mashian2016} at $z\in\{7.9, 6.8, 4.9\}$. The \citet{Mashian2016} curves are corrected to assume a \citet{Chabrier2003} IMF using the calibrations proposed by \citet{KE2012}. Tabulated quantities of the \bluetides\ predicted are given in Table \ref{tab:SFRDF}.}
\label{fig:SFRDF}
\end{figure}

Another fundamental description of galaxy population is the star formation rate (SFR) distribution function (SFR-DF). \bluetides\ predictions for the SFR-DF are shown, alongside observational constraints at $z\in\{4.9,6.8,7.9\}$ from \citet{Mashian2016} in Fig. \ref{fig:SFRDF}. The general shape of the predicted SFR-DF is similar to the galaxy stellar mass function and similarly lacks a strong break. However, the SFRDF also evolves more slowly than the galaxy stellar mass function. The \citet{Mashian2016} $z\approx 7.9$ distribution function has both a higher normalisation at low-SFRs and contains fewer high-SFR galaxies. The lack of high-SFR galaxies may again suggest a modelling issue though may also reflect an observational bias. This is discussed in more depth in \S\ref{sec:photometric.modelling.dust} where we discuss predictions for dust attenuation. 

\subsection{Star Formation Histories}\label{sec:physical.SFH}

\begin{figure}
\centering
\includegraphics[width=20pc]{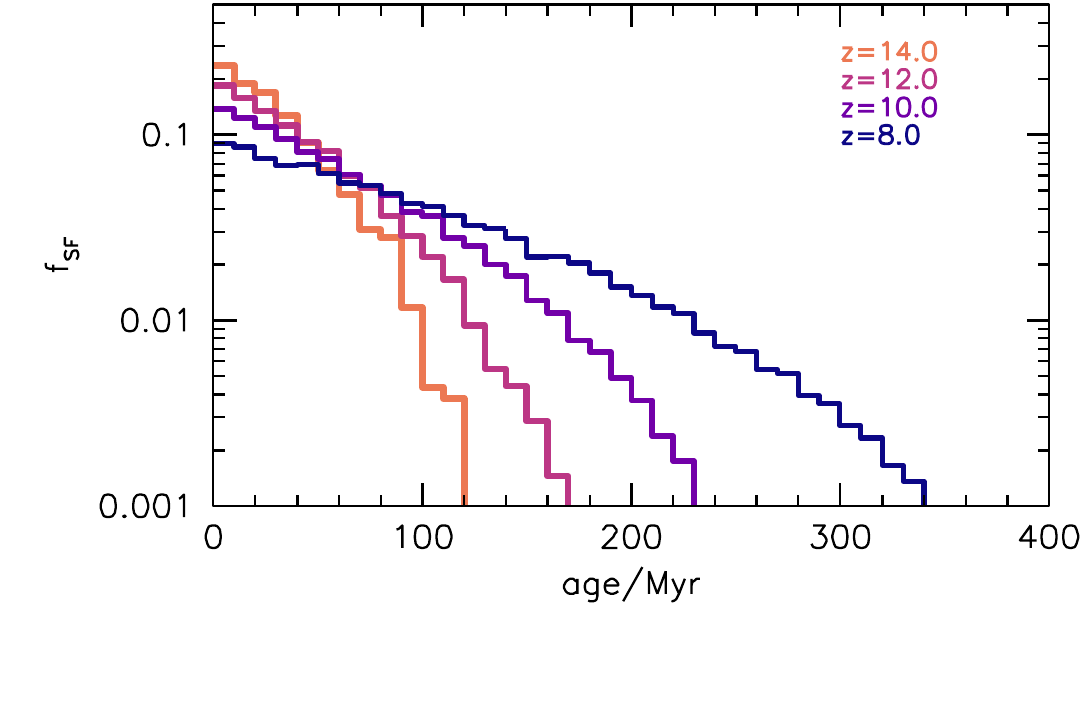}
\caption{The average star formation histories of galaxies with $M_{*}>10^{8}\,{\rm M_{\odot}}$ at $z\in\{14,12,10,8\}$. The figure shows the fraction of the total star formation occurring in each $\Delta t=10\,{\rm Myr}$ age-bin.}
\label{fig:SFH}
\end{figure}

\begin{figure}
\centering
\includegraphics[width=20pc]{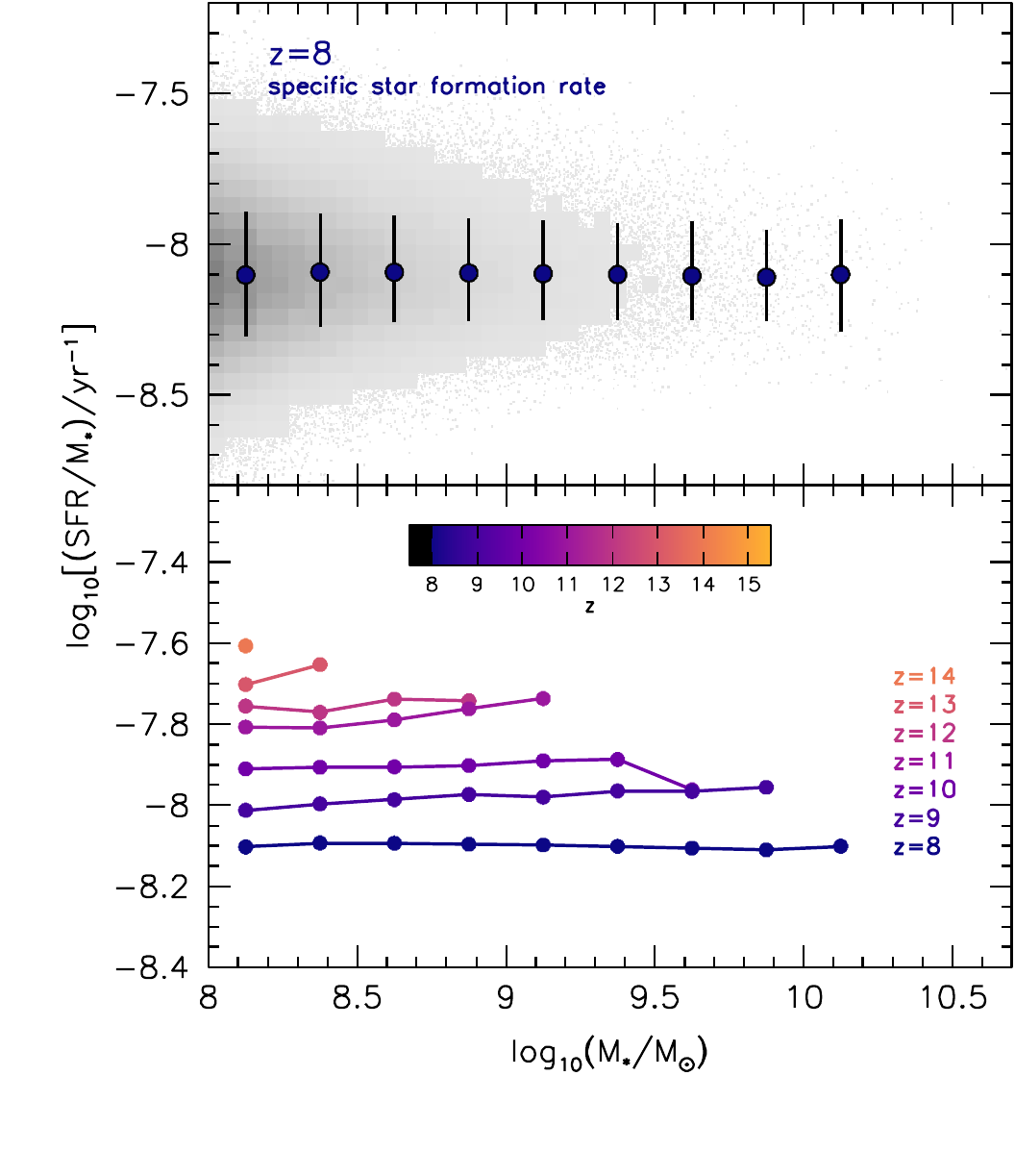}
\caption{The relationship between the specific star formation rate (${\rm SFR}/M_{*}$) and stellar mass predicted for galaxies at $z\in\{14,13,12,11,10,9,8\}$ by \bluetides. The top panel demonstrates the full distribution of sources at $z=8$ with the points denoting the median and the error-bars showing the central $68\%$ range. The lower panel shows only the median values for bins containing $>10$ galaxies at $z\in\{14,13,12,11,10,9,8\}$. The median specific star formation rates in stellar mass bins predicted by \bluetides\ are tabulated in Table \ref{tab:physical}.}
\label{fig:M_sSFR}
\end{figure}

\begin{figure}
\centering
\includegraphics[width=20pc]{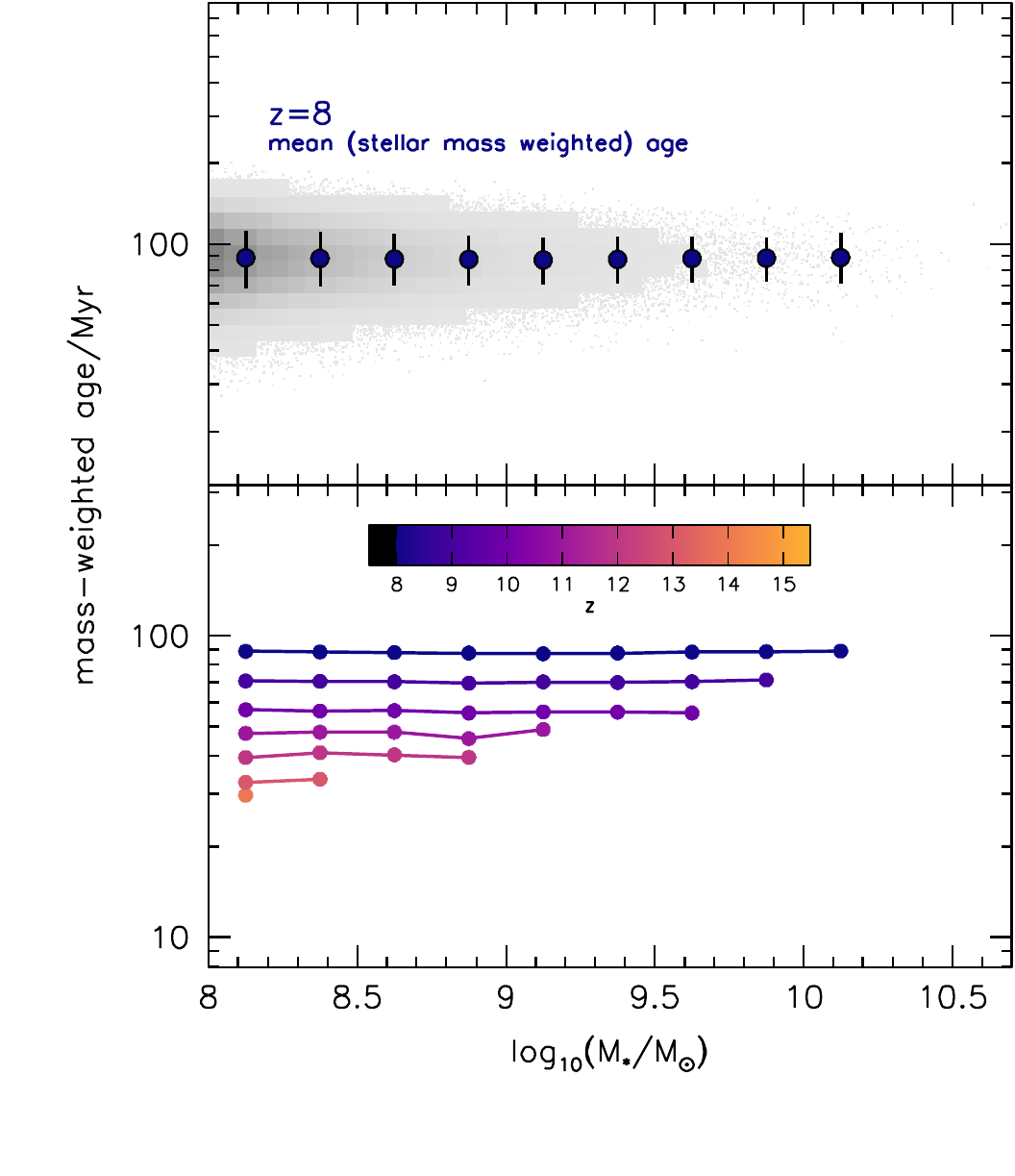}
\caption{The relationship between the mean stellar age and stellar mass predicted for galaxies at $z\in\{14,13,12,11,10,9,8\}$ by \bluetides. The top panel demonstrates the full distribution of sources at $z=8$ with the points denoting the median and the error-bars showing the central $68\%$ range. The lower panel shows only the median values for bins containing $>10$ galaxies at $z\in\{14,13,12,11,10,9,8\}$. The median ages in stellar mass bins predicted by \bluetides\ are tabulated in Table \ref{tab:physical}.}
\label{fig:ages}
\end{figure}

At all the redshifts simulated by \bluetides\ the average star formation activity in galaxies is increasing rapidly, though the rate of this increase slows at later times. The average star formation histories of galaxies with stellar masses $>10^{8}\,{\rm M_{\odot}}$ are shown in Fig. \ref{fig:SFH}. Within the range probed by \bluetides\ there is little variation in the shape of the star formation history with stellar mass. This can also be seen in Figs. \ref{fig:M_sSFR} and \ref{fig:ages} where we show the average specific star formation and mean stellar ages in different mass bins. Both quantities show no correlation with stellar mass over the range which we are sensitive suggesting that star formation has not yet been quenched in these systems. The lack of quenching in our simulated galaxies is not entirely surprising as the mass range does not yet encompass many galaxies with $M_{h}>10^{12}\,{\rm M_{\odot}}$ where inflows, and thus star formation, is expected to be suppressed \citep[e.g.][]{Finlator2011}. It is worth noting however there is a tentative indication of some suppression in the most massive halos, however at $z=8$ there are not yet enough to have a clear picture.

While there is no correlation with stellar mass both the average specific star formation rate and mean stellar age evolve strongly with redshift. For example, from $z=14\to 8$ average mass-weighted stellar ages increase from approximately $30\to 90\,{\rm Myr}$ while specific star formation rates drop by around $0.5\,{\rm dex}$.

\subsection{Metal Enrichment}\label{sec:physical.Z}

\begin{figure}
\centering
\includegraphics[width=20pc]{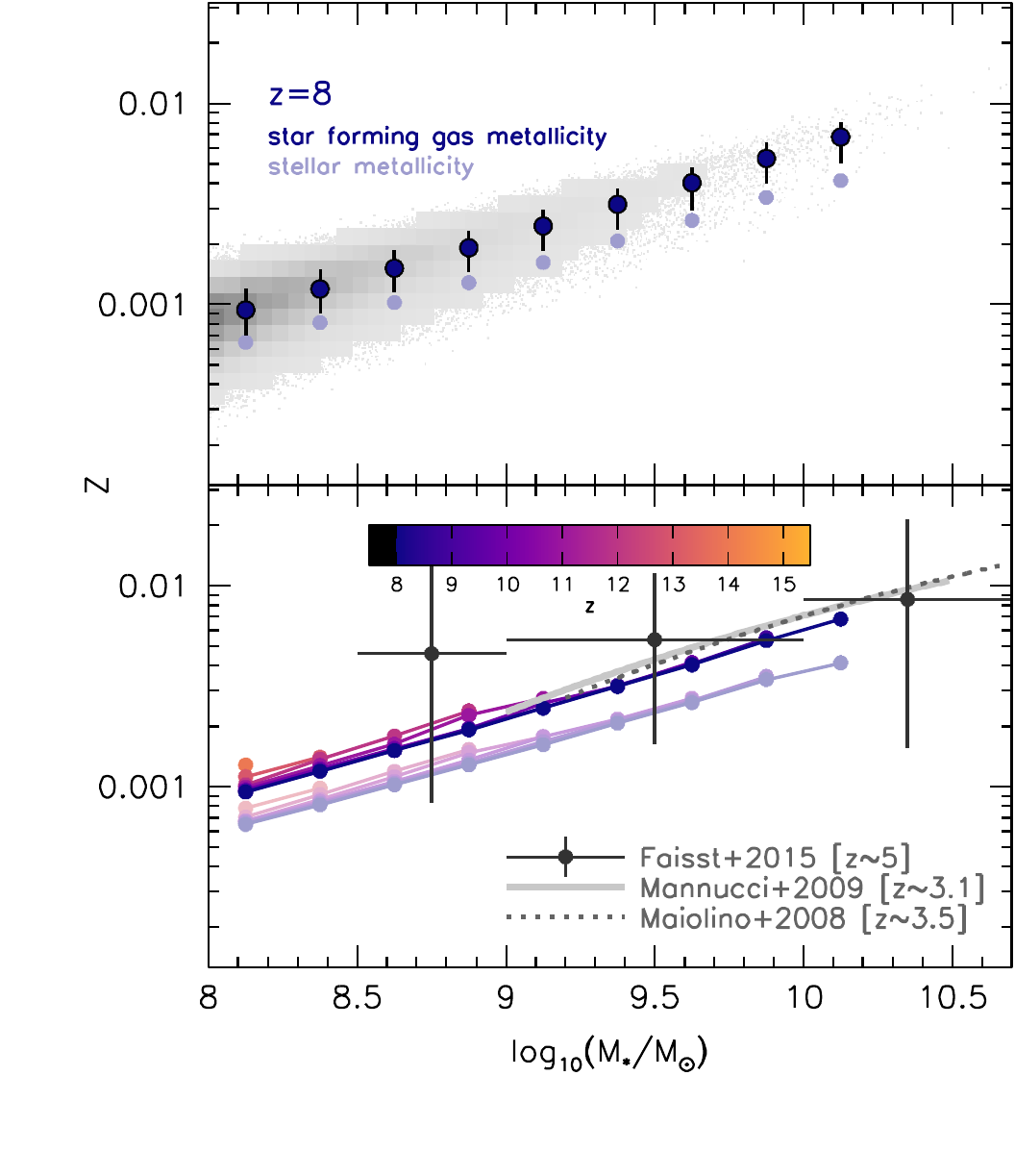}
\caption{The stellar (light points) and star forming gas (dark points) metallicities of galaxies in \bluetides. The 2D histogram in the top panel shows all objects with $M>10^{8}\,{\rm M_{\odot}}$ at $z=8$. Points denoting the median and central $68\%$. The lower panel shows only the median values for bins containing $>10$ galaxies at $z\in\{14,13,12,11,10,9,8\}$. Observational constraints from \citet{Maiolino2008}, \citet{Mannucci2009}, and \citet{Faisst2016} at $z\sim 3.5$, $z\sim 3.1$, and $z\sim 5$ respectively are also shown. Observational measurements of the stellar mass assume a \citet{Chabrier2003} initial mass function and metallicities were converted to a mass-fraction assuming $12+\log_{10}(O/H)_{\odot} = 8.69$ and $Z_{\odot}=0.02$. The median metallicities in stellar mass bins predicted by \bluetides\ are tabulated in Table \ref{tab:physical}.}
\label{fig:metallicities}
\end{figure}

As galaxies assemble stellar mass in the simulation the average metallicity of both the gas and stars increases. This can be seen in Fig. \ref{fig:metallicities} where we show both the average mass-weighted stellar and star forming gas phase metallicity as a function of stellar mass. The trend of metallicities with stellar masses increases as ${\rm d}\log_{10}Z/{\rm d}\log_{10}M_*\approx 0.4$. This trend is similar to observational measurements, using rest-frame optical strong line diagnostics, from \citet{Maiolino2008} (at $z\approx 3.5$) and \citet{Mannucci2009} (at $z\approx 3.1$). The normalisation of the simulated mass-metallicity relationship at $z\approx 8$ is also similar to that found at $z\sim 3$ by \citet{Maiolino2008} and \citet{Mannucci2009} using rest-frame optical diagnostics and at $z\sim 5$ by \citet{Faisst2016} using rest-UV absorption complexes.

\section{Photometric Properties}\label{sec:photometric}

\subsection{Modelling Galaxy Photometry}\label{sec:photometric.modelling}

We build up the spectral energy distribution (SED) of each galaxy on a star particle by star particle basis. Firstly, we assign a pure stellar SED to each particle on the basis of its mass, age, and chemical composition. We adopt the {\sc Pegase.2} \citep{pegase} stellar population synthesis (SPS) model combined with a \citet{Chabrier2003} initial mass function (IMF) over $0.1-100\,{\rm M_{\odot}}$. The emission from each star particle is then modified to take into account reprocessing by both dust and gas as described below. 

\subsubsection{Nebular Continuum and Line Emission Modelling}\label{sec:photometric.modelling.nebular}

We use the {\sc cloudy} photoionisation code to model the effect of reprocessing by H{\sc ii} surrounding stars. The hydrogen density is chosen to be $100\,{\rm cm^{-3}}$ and the chemical composition of the gas is set to the metallicity of the star particle scaled by solar abundances. We assume a uniform covering fraction of $0.85$ thereby leaving sufficient LyC photons to reionise the Universe. 

The implications of the choice of SPS model, initial mass function, and Lyman continuum (LyC) escape fraction on the spectral energy distributions are discussed in more detail in \citet{Wilkins2016b} and \citet{Wilkins2016c}. While these assumptions can result in large systematic effects, the effect on the rest frame far-UV ($150\,{\rm nm}$) is relatively small as nebular emission contributes only around $10\%$ of the total luminosity and variations due to the choice of model typically changing luminosities by $<0.1\,{\rm dex}$ \citep{Wilkins2016c}.

\subsubsection{Dust Attenuation}\label{sec:photometric.modelling.dust}

To estimate the dust attenuation in \bluetides\ we employ a scheme which links the metal density integrated along parallel lines of sight to the dust optical depth $\tau$. 

In this model the rest-frame $V$-band ($0.55\,{\rm \mu m}$) dust optical depth ($\tau_{V}(x, y, z)$) is,
\begin{equation}
\tau_{V}(x, y, z) = \kappa \Sigma (x, y, z) = \int_{z'=0}^{z} \kappa \rho_\mathrm{metal}(x, y, z')\,{\rm d}z',
\end{equation}
where $\rho_\mathrm{metal}(x, y, z')$ is the metal density, and we have chosen the $z$ direction to be the line of sight direction. $\kappa$ is a normalization factor, a free parameter that is tuned to match the model with the observed $z\approx 8$ luminosity function (see \S\ref{sec:photometric.UVLF}).

First, the metal mass is painted to a 3-dimensional image with resolution of $0.2 h^{-1}\,{\rm ckpc}$. The image is passed through a Gaussian smoothing filter with a width of $r_s = 0.5 h^{-1}\,{\rm ckpc}$, the most probable smoothing length of gas particles that have collapsed into galaxies in the simulation. The parameter $r_s$ is also degenerate with $\kappa$. Secondly, we compute the cumulative sum of the image along the line of sight direction ($z$). After this procedure, the image contains the surface density of metals ($\Sigma(x, y, z)$) that contributes to the attenuation at any spatial location. Finally, we read off the values from the image at the location of each star particle. 

We employ an individual stellar cluster (ISC) approximation in the implementation. The star clusters are identified with a Friends-of-Friends algorithm with a linking length of $l = 2.0 h^{-1}\,{\rm ckpc}$. For each star cluster, we perform the above calculation for metal mass in the bounding box of the star cluster with a buffer region of $b=2.0 h^{-1}\,{\rm ckpc}$. We tested that the approximation is stable to reasonable changes in the linking length $l$ or the size of the buffer region. The ISC approximation allows us to focus the computational resource to locations in the simulation where the the dust attenuation is most relevant. At the high-redshifts ($z \ge 8$) simulated by \bluetides\ the ISC approximation provides a significant computational advantage comparing to a full volume ray tracing approach. At such high redshift, the attenuation due to chance-aligned galaxies can be neglected because the abundance of galaxies with very high metallicities is low. 

The optical depth at an arbitrary wavelength $\lambda$ is related to the $V$-band optical depth through an attenuation curve. We parameterise the attenuation curve as a power-law with index $\gamma$,
\begin{equation}
\tau_\lambda = \tau_V\times\left(\frac{\lambda}{0.55\,{\rm\mu m}}\right)^{\gamma}.
\end{equation}
For $\gamma$ we choose a value of $-1$ yielding an attenuation curve slightly flatter in the UV than the Pei et al. (1992) Small Magellanic Cloud  curve, but not as flat as the \citet{Calzetti2000} ``Starburst'' curve. 

\begin{figure}
\centering
\includegraphics[width=20pc]{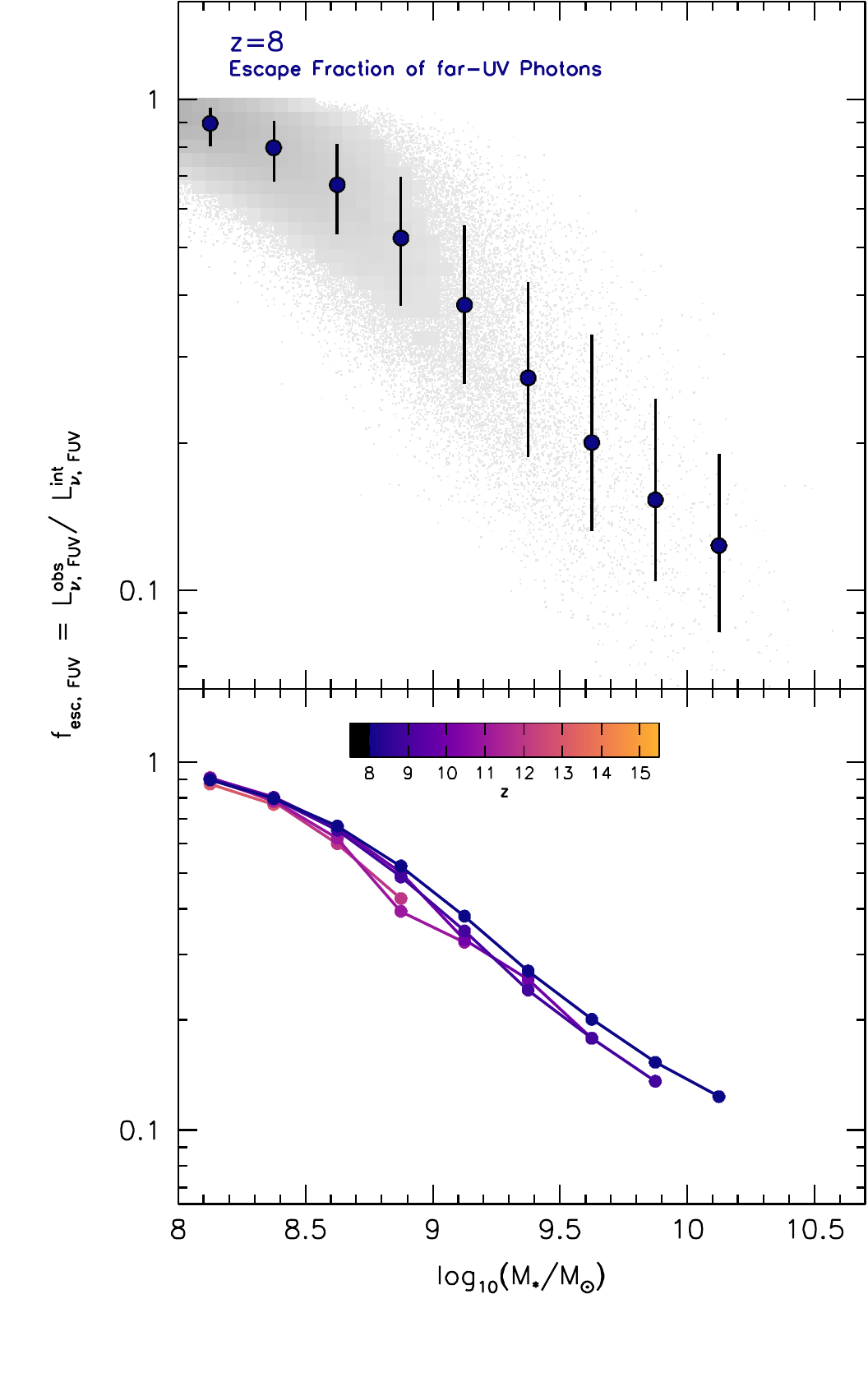}
\caption{The effective escape fraction of far-UV ($150\,{\rm nm}$) photons as a function of stellar mass. The median far-UV escape fractions in stellar mass bins predicted by \bluetides\ are tabulated in Table \ref{tab:L_fesc}.}
\label{fig:L_fesc}
\end{figure}

The predicted surface density of metals is strongly correlated with the stellar mass and intrinsic luminosity. This results in a strong trend of the average UV attenuation with both the stellar mass and intrinsic UV luminosity, albeit with considerable scatter (see Fig. \ref{fig:L_fesc}). At a fixed stellar mass the attenuation is predicted to decrease slightly to higher redshift. 

However, the formation of dust and metals, while linked to some degree, are not expected to trace one another exactly \citep[see modelling by][]{Mancini2015}. Consequently, such a simple model is unlikely to fully capture the redshift and luminosity dependence of dust attenuation, especially at the highest redshifts where the formation of dust in AGB stars or in-situ in the ISM has not had time to occur. This may then suggest that our dust model produces too much attenuation at the highest redshifts. Indeed, this is perhaps hinted at by the recent discovery \citep{Oesch2016} of an exceptionally bright ($M\approx -22$) and blue (and therefore likely dust-poor) galaxy at $z\approx 11$. While the discovery of this object is consistent with predictions based on intrinsic luminosities \citep{Waters2016a} it would be very unexpected based on dust attenuated luminosities predicted using our model. Given the uncertainties introduced by the dust model, particularly at $z>10$, throughout this work we consider predictions based on both the intrinsic luminosities and the dust attenuated luminosities.

A consequence of the desire to fit observations of the $z\sim 8$ far-UV luminosity function is the prediction that there exist a number of massive, heavily dust-obscured galaxies. The existence of these galaxies, which would not appear in Lyman-break selected samples, explains the discrepancy between predictions from \bluetides\ and current observational constraints on the galaxy stellar mass function and star formation rate distribution function (see \S\ref{sec:physical.GSMF} and \S\ref{sec:physical.SFRDF}). Unfortunately, the relative faintness and rarity of these objects means they are unlikely to be identified in current IR/sub-mm observations. However, massive heavily obscured intensely star forming galaxies have been identified at lower redshift \citep[e.g. HFLS3 at $z=6.34$:][]{Riechers2013} suggesting that such objects can and do exist in the relatively early Universe.

\subsection{Spectral Energy Distributions}\label{sec:photometric.SED}

\begin{figure*}
\centering
\includegraphics[width=40pc]{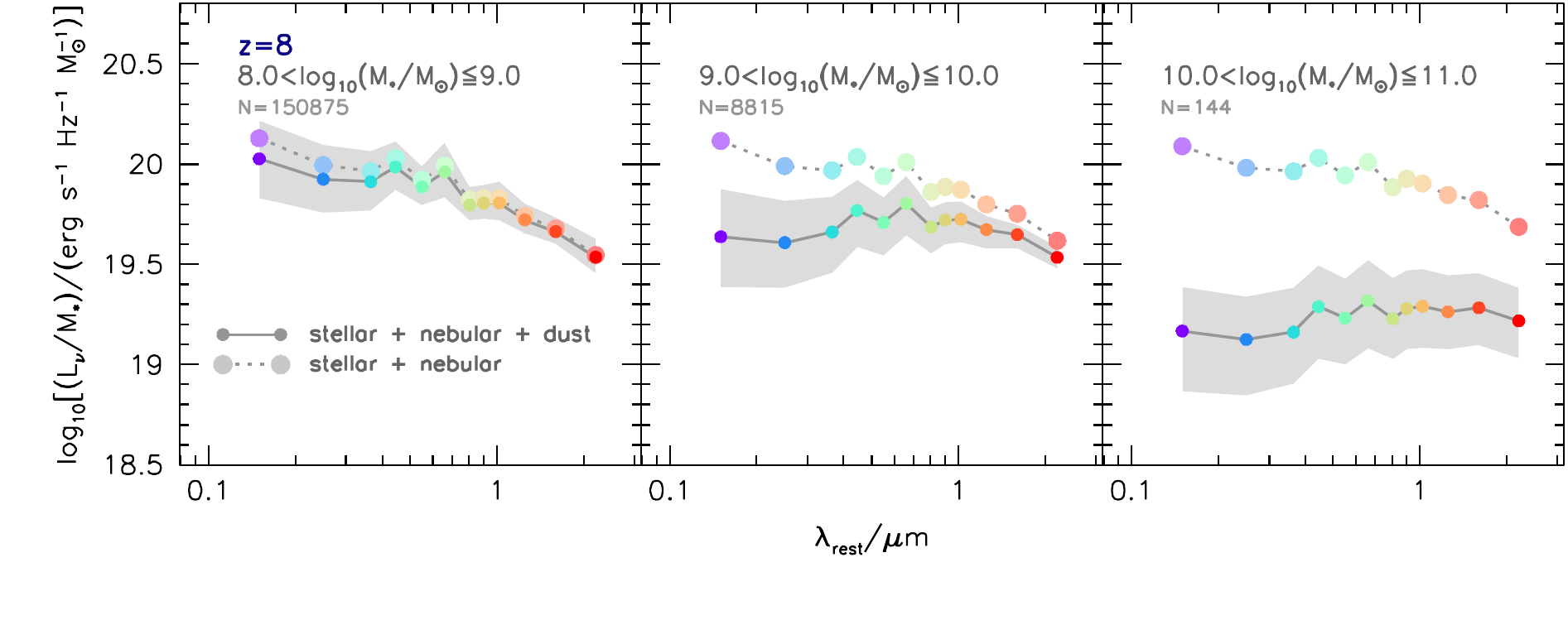}
\caption{The average observed and unattenuated SEDs (expressed per unit stellar mass) in three mass bins at $z=8$.}
\label{fig:SED_M}
\end{figure*}

The resulting average intrinsic (including nebular continuum and line emission) and observed specific\footnote{That is, expressed per unit stellar mass.} spectral energy distributions are shown, for three mass bins at $z=8$, in Fig. \ref{fig:SED_M}. 

The average intrinsic SEDs are generally very blue, reflecting the ongoing star formation activity, young ages, and low metallicities in the sample. While the shape of the SEDs in each mass bin is very similar, the most massive galaxies have slightly redder SEDs reflecting the higher metallicity of the stellar populations. A more detailed analysis of the pure stellar and intrinsic SEDs is contained in \citet{Wilkins2016c}. 

\begin{figure}
\centering
\includegraphics[width=20pc]{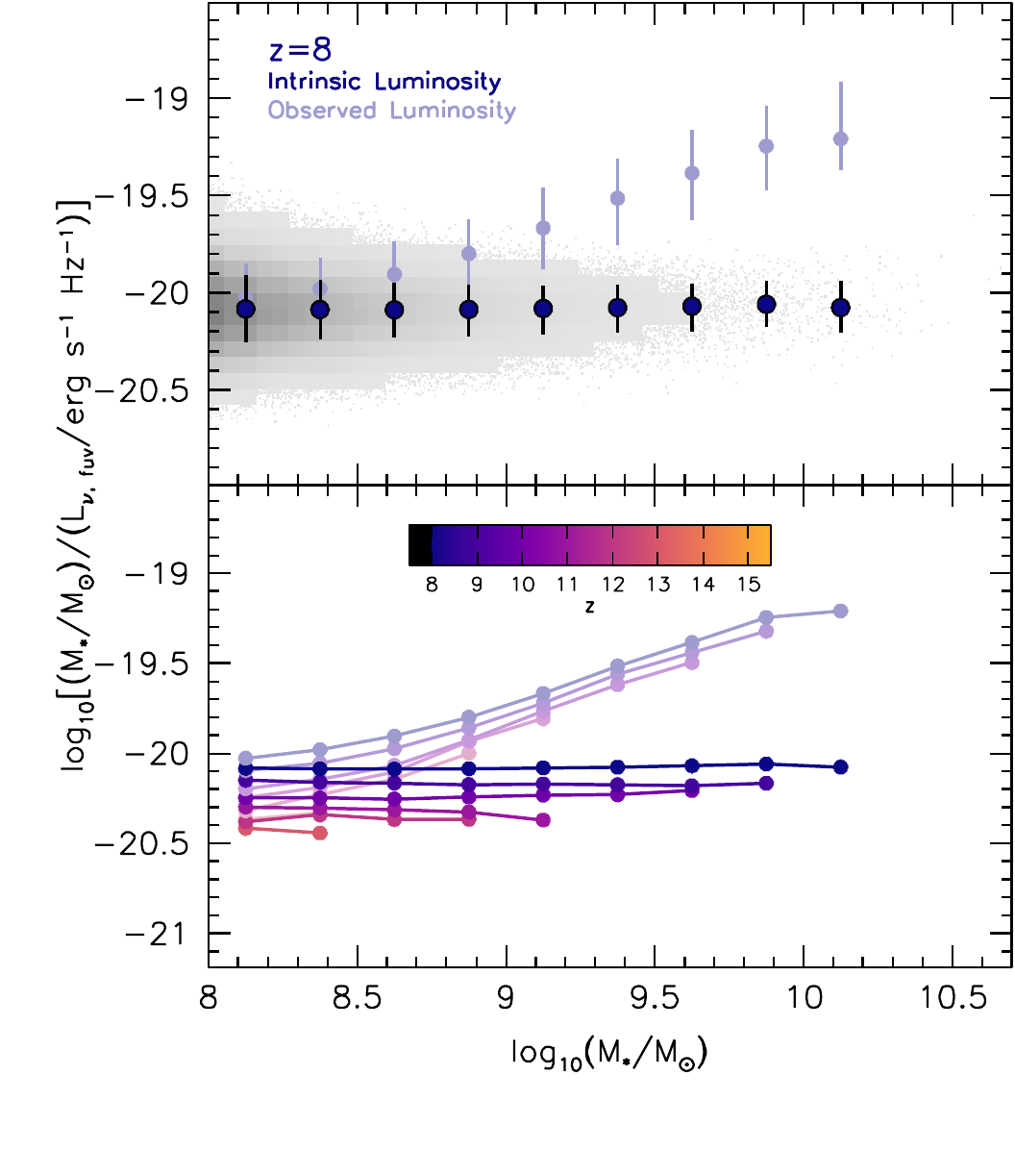}
\caption{The intrinsic and dust attenuated far-UV mass-to-light ratios as function of stellar mass and redshift. The median intrinsic and observed far-UV mass-to-light ratios in stellar mass bins predicted by \bluetides\ are tabulated in Table \ref{tab:MTOL}.}
\label{fig:MTOL}
\end{figure}

As noted in the previous section, the most massive galaxies also suffer much higher attenuation due to dust resulting in redder observed SEDs and higher mass-to-light ratios. The trend of higher mass-to-light ratios at higher stellar mass can be seen more clearly in Fig. \ref{fig:MTOL}. Fig. \ref{fig:MTOL} also shows the evolution with redshift demonstrating that stellar mass-to-light ratios increase to lower redshift. This predominantly reflects the increasing age of the stellar populations to lower redshift.

\subsection{Luminosity Functions}\label{sec:photometric.UVLF}

\begin{figure*}
\centering
\includegraphics[width=40pc]{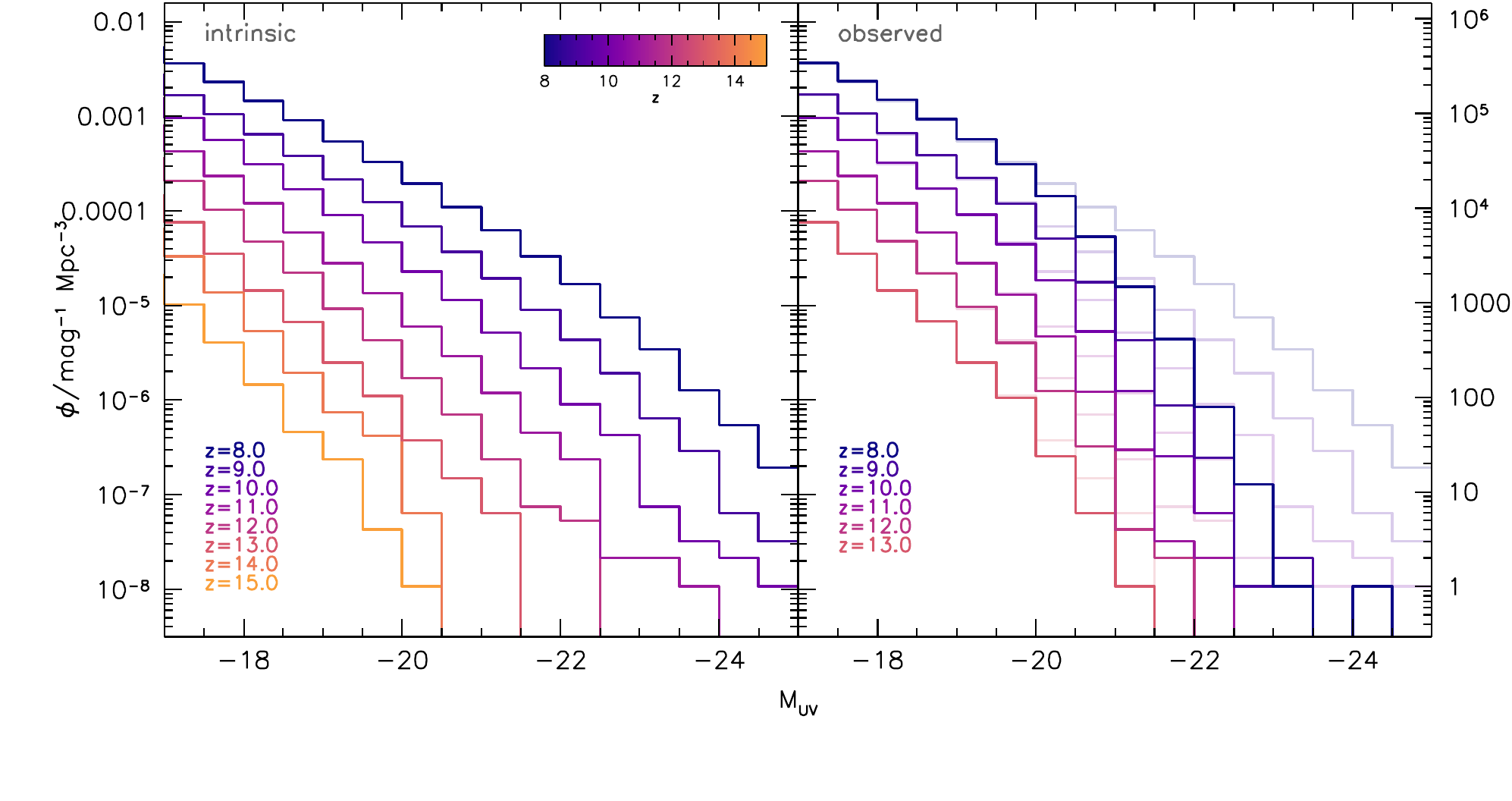}
\caption{Intrinsic (left panel) and dust attenuated (observed, right panel) rest-frame far-UV ($150\,{\rm nm}$) luminosity functions. Observations at $z\approx 8$ and $10.4$ from Bouwens et al.\ (2015) are shown for comparison. The scale of the right-hand axis shows the number of galaxies in each bin magnitude in the simulation. Tabulated quantities of the \bluetides\ predicted are given in Table \ref{tab:UVLF}.}
\label{fig:UVLF}
\end{figure*}

\begin{figure*}
\centering
\includegraphics[width=40pc]{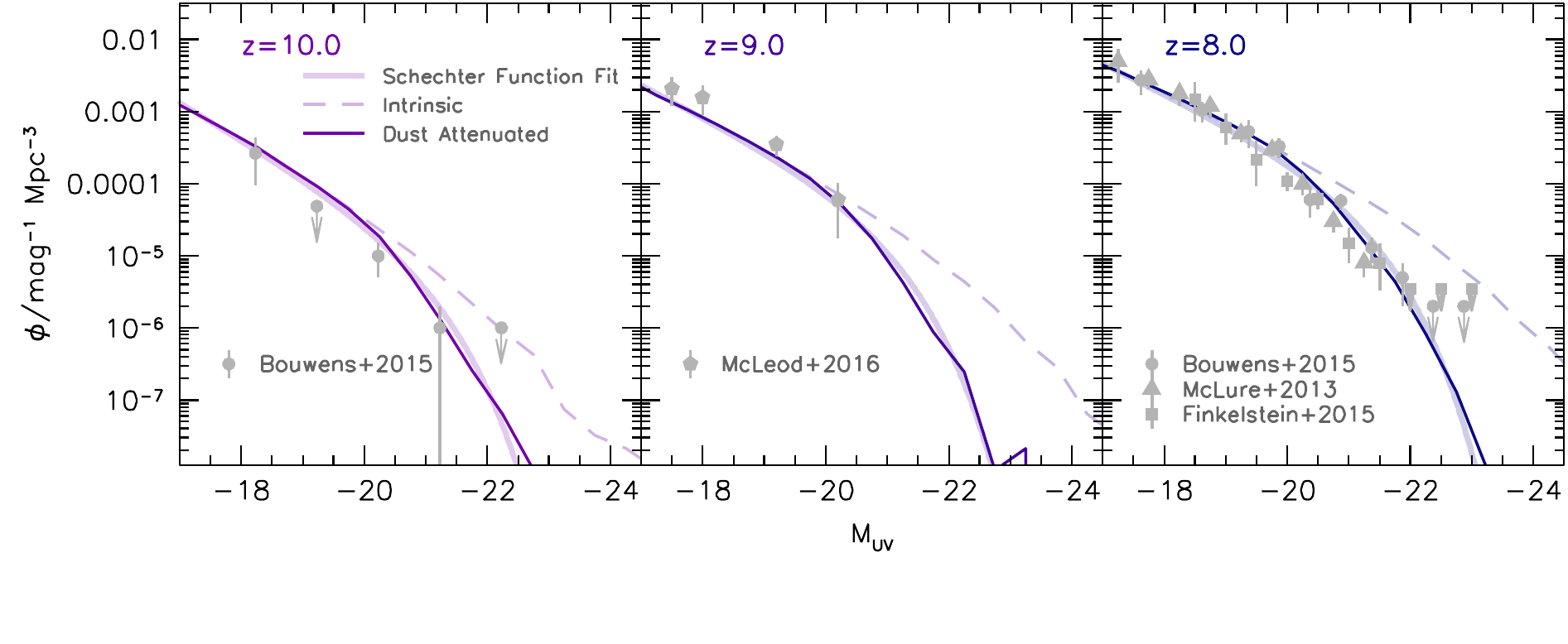}
\caption{Intrinsic (left panel) and dust attenuated (observed, right panel) rest-frame far-UV ($150\,{\rm nm}$) luminosity functions. Observations at $z\approx 8$ and $10.4$ from Bouwens et al.\ (2015) are shown for comparison. The scale of the right-hand axis shows the number of galaxies in each bin magnitude in the simulation. Tabulated quantities of the \bluetides\ predicted are given in Table \ref{tab:UVLF}.}
\label{fig:UVLF_multi}
\end{figure*}

The luminosity function (LF) is an incredibly useful statistical description of the galaxy population. In Fig. \ref{fig:UVLF} we present both the intrinsic and dust attenuated far-UV luminosity functions at $z=8\to 15$. In Fig. \ref{fig:UVLF_multi} we show the intrinsic and attenuated UV LFs at $z\in\{8,9,10\}$ together with current observational constraints. Both the intrinsic and observed luminosity functions demonstrate the rapid expected build up of the galaxy population at high-redshift. For example, the number of $M=-19$ objects increases by a factor of around 1000 from $z=15\to 8$. The rapid decline of the LF to high-redshift poses challenges for the observational identification of galaxy populations at $z>12$ even using \jwst. This is explored in more detail in Wilkins et al. {\em submitted} where we make predictions for the surface density of sources at $z>8$ including the effects of field-to-field, or cosmic, variance.

The observed LF is generally similar to the intrinsic LF at faint luminosities ($M>-20$). At brighter luminosities there is stronger steepening of the LF reflecting the increasing strength of dust attenuation. As noted earlier our dust model is tuned to match the $z\approx 8$ observed UV LF. However, it is important to stress that this only makes a significant difference at relatively bright luminosities ($M<-20$); at fainter luminosities there simply is not the surface density of metals (and therefore inferred dust) to yield significant attenuation. The excellent fit at fainter luminosities is then simply a consequence of the physics employed in the model and not a resulting of tuning using the dust model. However, while the faint end of the LF is unaffected by our choice of dust model it can be systematically affected by the choice of initial mass function (and to a lesser extent choice of SPS model); see \citet{Wilkins2016c}. Adopting an IMF yielding more low-mass stars than our assumed IMF \citep[e.g. a pure][IMF extended down to $0.1\,{\rm M_{\odot}}$]{Salpeter1955} would uniformly reduce the luminosities of our galaxies, shifting the LF to fainter luminosities.

\begin{figure}
\centering
\includegraphics[width=20pc]{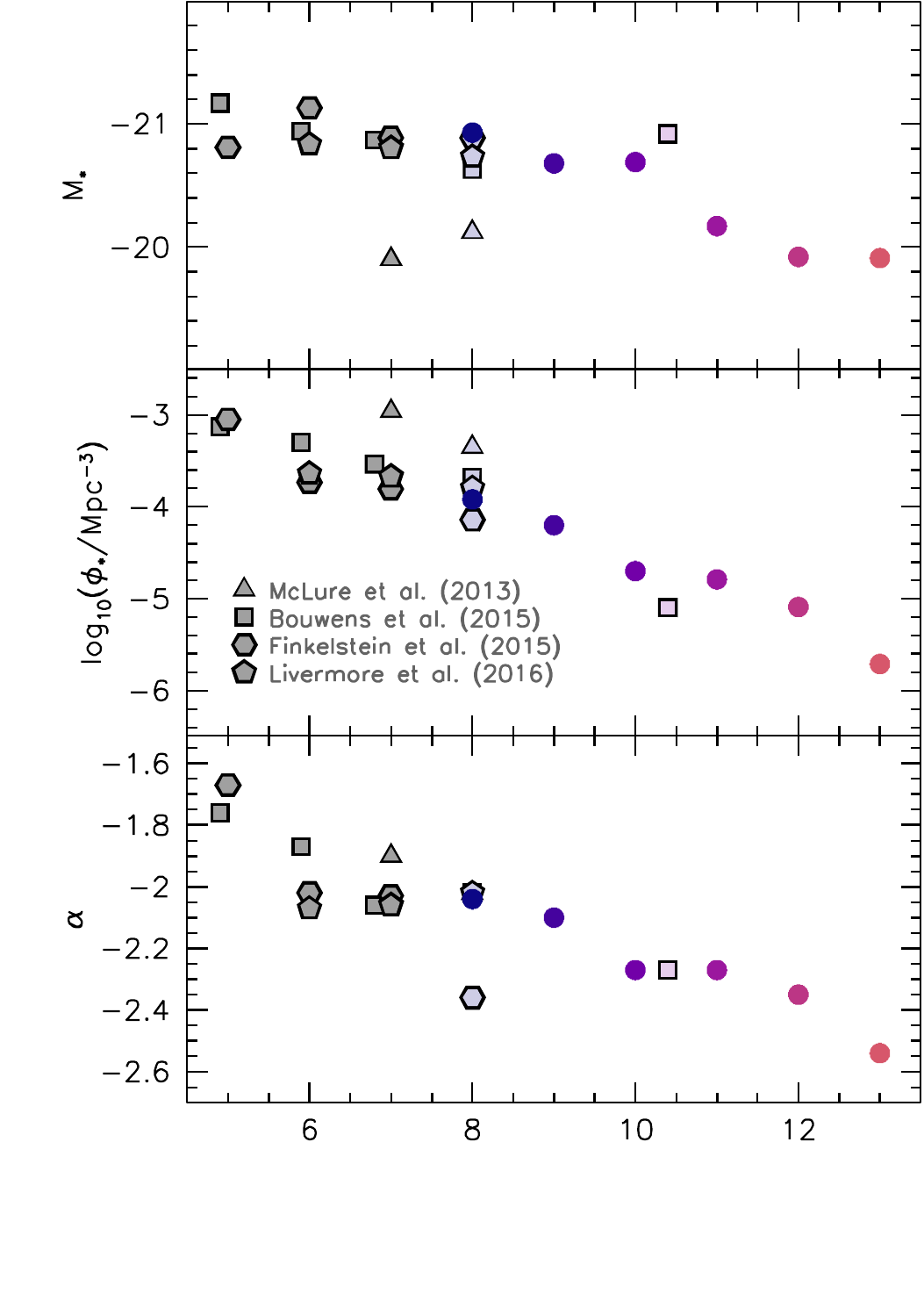}
\caption{Redshift evolution of the best fit Schechter function parameters of the simulated observed UV LF.}
\label{fig:parameters_redshift}
\end{figure}

We also fit the dust attenuated far-UV LF by a Schechter function and find that the function provides a good overall fit to shape of the LF, as shown in Fig. \ref{fig:UVLF_multi} at $z\in\{8,9,10\}$. The evolution of the Schechter function parameters is shown in Fig. \ref{fig:parameters_redshift} with the parameters listed in Table \ref{tab:parameters_redshift} alongside various observational constraints at $z=4-10$. All three parameters decrease to higher redshift and overlap with observational constraints (and extrapolations from lower-redshift).

\begin{table}
\caption{Best fit Schechter function parameters for the observed UV luminosity function.}
\label{tab:parameters_redshift}
\begin{tabular}{cccc}
\hline
$z$ & $M^{*}$ & $\log_{10}(\phi^{*}/{\rm Mpc^{-3}})$ & $\alpha$ \\
\hline
13 & -19.91 & -5.71 & -2.54\\
12 & -19.92 & -5.09 & -2.35\\
11 & -20.17 & -4.79 & -2.27\\
10 & -20.69 & -4.70 & -2.27\\
9 & -20.68 & -4.20 & -2.10\\
8 & -20.93 & -3.92 & -2.04\\
\hline
\end{tabular}
\end{table}

\section{Super-massive black-holes}\label{sec:SMBHs}

\begin{figure}
\centering
\includegraphics[width=20pc]{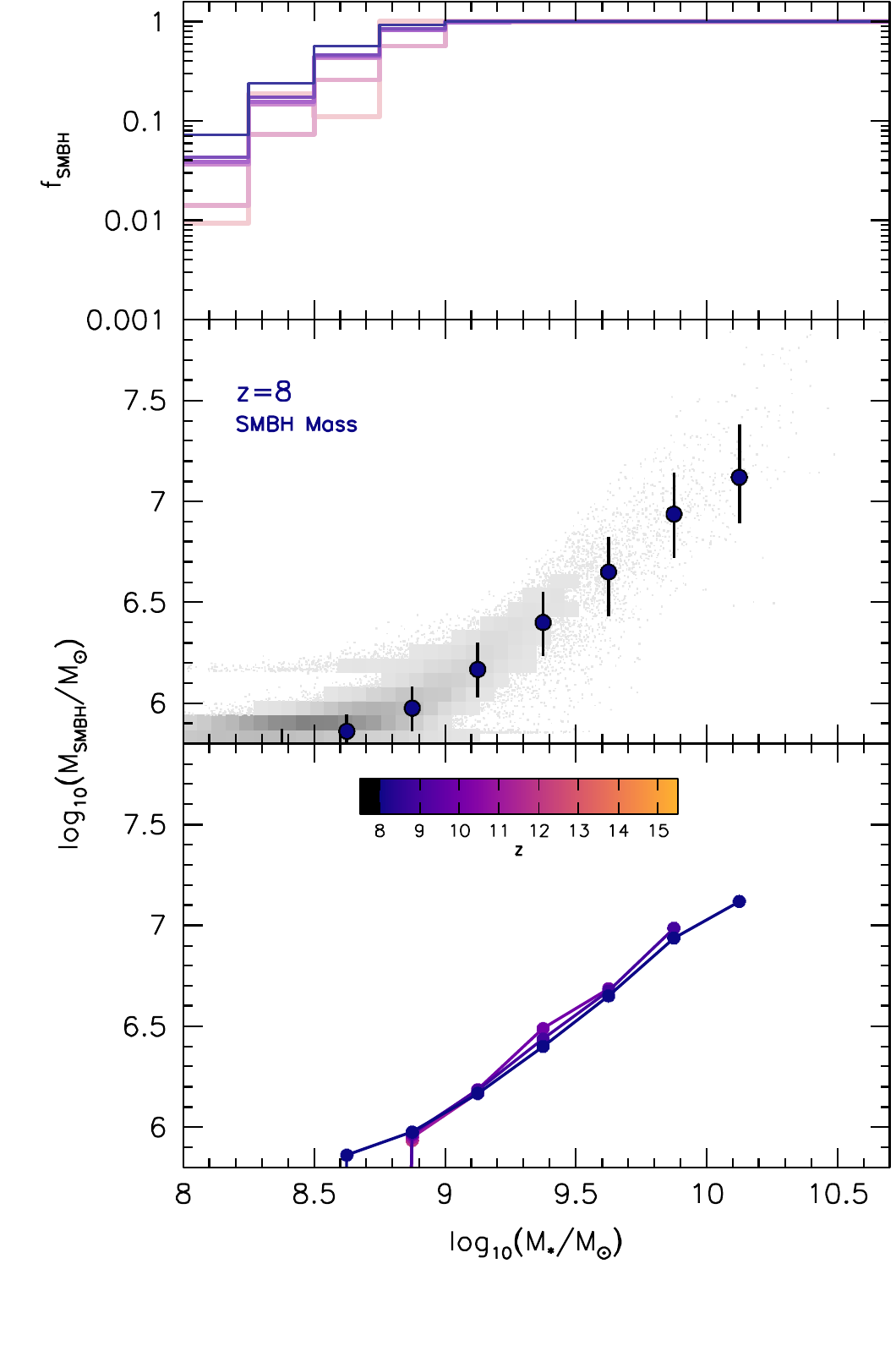}
\caption{{\em Top-panel} The fraction of galaxies in halos hosting a SMBH. {\em Middle/bottom-panels} The relationship between the mass of the central super-massive black hole and the stellar mass of galaxies in \bluetides. The top panel demonstrates the full distribution of sources at $z=8$ with the points denoting the median and the error-bars showing the central $68\%$ range within uniform stellar mass bins. The lower panel shows only the median of stellar mass bins for $z\in\{14,13,12,11,10,9,8\}$. Median SMBH masses in stellar mass bins are tabulated in Table \ref{tab:SMBH}.}
\label{fig:M_MSMBH}
\end{figure}

Super-massive black-holes (SMBHs) are incorporated into \bluetides\ by first seeding halos with $5\times 10^5\,h^{-1}\,{\rm M_{\odot}}$ mass black holes once they reach a dark matter mass greater than $>5\times 10^{10}\,h^{-1}\,{\rm M_{\odot}}$. Fig. \ref{fig:M_MSMBH} shows both the fraction of galaxies hosting a SMBH and the SMBH mass as a function of stellar mass. The majority of galaxies with stellar masses below $\sim 10^{8.5}\,{\rm M_{\odot}}$ occupy halos that have yet to be seeded with a SMBH while virtually all above $\sim 10^{9}\,{\rm M_{\odot}}$ are in halos containing a SMBH. At higher stellar masses the SMBH and stellar mass are strongly correlated, albeit with significant scatter. The formation and evolution of super-massive black-holes in \bluetides\ is discussed in more detail in Di Matteo et al. {\em in-prep}.

\subsection{Contribution to far-UV luminosities}\label{sec:SMBHs.UV}

The rate at which mass is accreted from the halo onto the SMBH ${\rm d}M_{\bullet}/{\rm d}t$ can be used to estimate the bolometric luminosity $L_{\rm bol}$, 
\begin{equation}
L_{\rm bol} = \eta \frac{{\rm d}M_{\bullet}c^2}{{\rm d}t}
\end{equation}
where $\eta$ is the efficiency and is assumed to be $0.1$. 

\begin{figure}
\centering
\includegraphics[width=20pc]{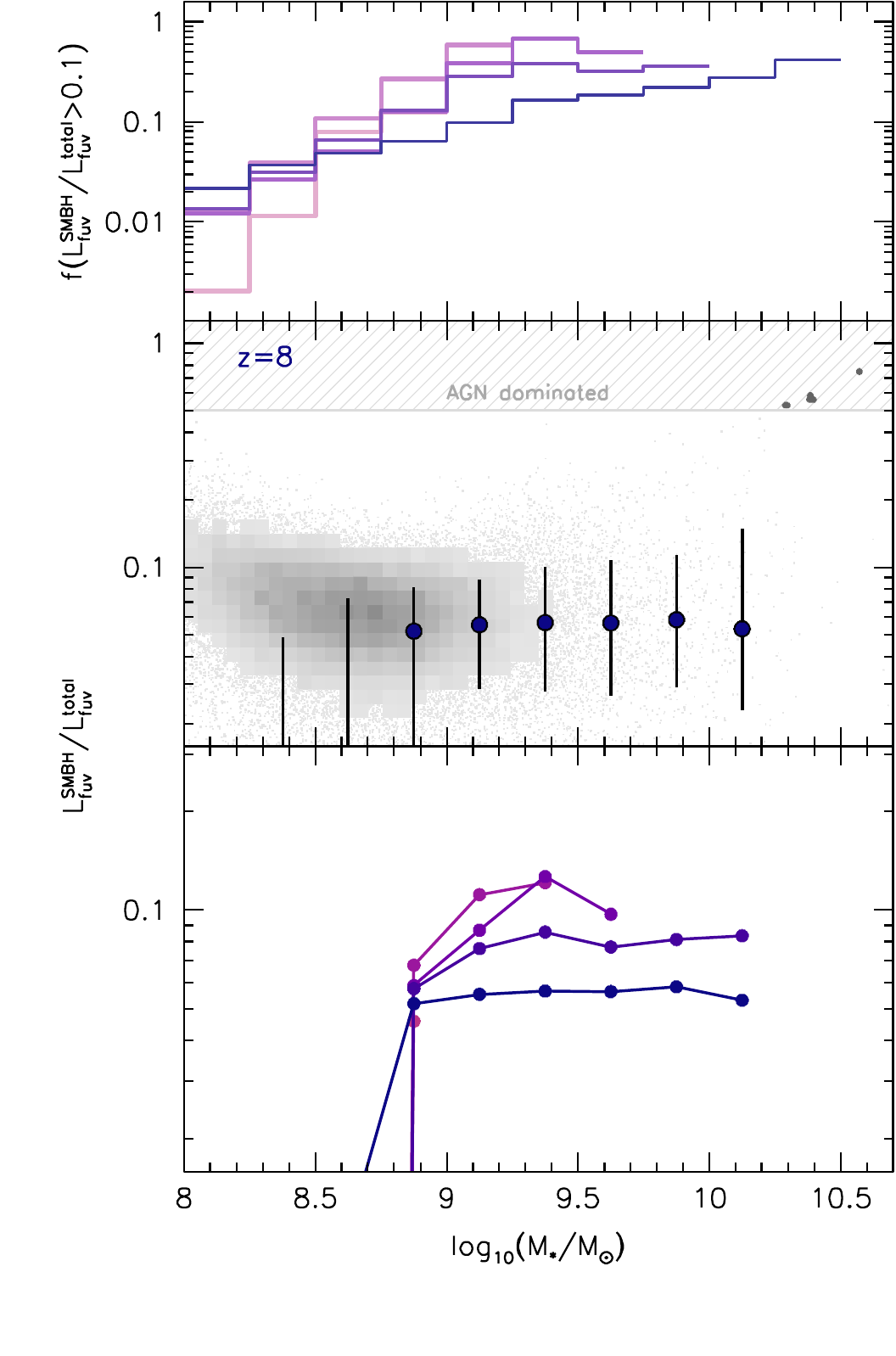}
\caption{{\em Top-panel} The fraction of galaxies in which the SMBH contributes $>10\%$ of the far-UV ($150\,{\rm nm}$) luminosity. {\em Middle/bottom-panels} The fractional contribution of SMBHs to the total far-UV luminosity as a function of stellar mass. Note: the statistics are calculated for all galaxies (i.e. including those galaxies which have not yet been seeded with a SMBH). The top panel shows the full distribution of sources at $z=8$ with the large points denoting the median and the error-bars showing the central $68\%$ range within uniform stellar mass bins. The small dark individual grey points denote objects where the SMBH accounts for $>50\%$ of the total far-UV luminosity. The lower panel shows only the median of stellar mass bins for $z\in\{11,10,9,8\}$.}
\label{fig:SMBH_LCont}
\end{figure}

Assuming a bolometric correction of $2.25$\footnote{A larger bolometric correction as suggested by \citet{Runnoe2012} would reduce the predicted luminosities of the SMBHs and thus the fractional contribution to the galaxy luminosity.} \citep{Fontanot2012} we estimate the far-UV luminosities of the SMBHs. In Figure \ref{fig:SMBH_LCont} we show the fractional contribution of the SMBH to the total far-UV luminosity. In galaxies hosting a SMBH the SMBH on average contributes only approximately $5\%$ of the total (intrinsic stellar + SMBH) far-UV luminosity. The fraction of galaxies hosting a SMBH that contributes $>10\%$ of the total far-UV luminosity increases with stellar mass. In galaxies at $z=8$ with stellar masses $>10^{10}\,{\rm M_{\odot}}$ approximately $25\%$ of galaxies host a SMBH which contributes $>10\%$ of the far-UV luminosity. In six of the most massive ($M_*=10^{10-10.6}\,{\rm M_{\odot}}$) galaxies the total far-UV luminosity is dominated by accretion on to the central SMBH.

\section{Conclusions}\label{sec:c}

We have used the large cosmological hydrodynamical simulation \bluetides\ to make detailed predictions of the formation and evolution of galaxies during the epoch of reionisation.

\begin{itemize}

\item The stellar to dark matter mass ratio increases by approximately $0.6\,{\rm dex}$ over the range $\log_{10}(M_{\rm DM}/M_{\odot})=10.5\to 12$. This increase is similar to that predicted by Moster et al. (2013) although with a different normalisation.

\item The galaxy stellar mass function evolves rapidly from $z=15\to 8$, with the number density of galaxies with stellar masses $\approx 10^{8}\,{\rm M_{\odot}}$ increasing by a factor of ten thousand. At higher-mass the increase is more rapid leading to an evolution in the shape of the mass function. The $z=8$ galaxy stellar mass function closely matches recent observational estimates from \citet{Song2015}.
  
\item On average the star formation histories of galaxies are rapidly increasing.

\item There is little dependence of the shape of the star formation history on the mass of galaxies. This is reflected in the lack of a correlation between the specific star formation rates and ages with stellar mass. Specific star formation rates and ages do however evolve strongly with redshift. For example, the mass-weighted age of galaxies at $z=8$ is approximately $90\,{\rm Myr}$ while at $z=12$ it is $\approx 40\,{\rm Myr}$.

\item Stellar and gas metallicities are strongly correlated with stellar mass, though evolve only weakly with redshift (at fixed stellar mass).

\item The average integrated surface density of metals is strongly correlated with stellar mass and intrinsic UV luminosity. Using a model in which the dust attenuation is linked to the surface density of metals suggests the far-UV ($150\,{\rm nm}$) escape fraction strongly anti-correlates with intrinsic luminosity (i.e. the intrinsically brightest objects are most heavily affected by dust attenuation). 

\item The predicted far-UV luminosity function closely matches observations at faint the end. At the bright end good agreement can be obtained by linking the integrated surface density of metals to the dust optical depth. The expected surface densities of sources suggest relatively few galaxies will be identified by the {\em James Webb Space Telescope} above $z=12$.

\item In galaxies hosting an AGN at $z>8$ the AGN is, on average, predicted to only make a small ($\approx 5\%$) contribution to the far-UV luminosity. The fraction of galaxies in which the AGN contributes $>10\%$ of the total far-UV luminosity increases with stellar mass; at stellar masses $>10^{10}\,{\rm M_{\odot}}$ approximately $25\%$ of galaxies host a SMBH which contributes $>10\%$ of the total far-UV luminosity. At $z=8$ we identify 6 galaxies (all amongst the most massive systems) within the \bluetides\ volume in which the far-UV is dominated by the SMBH.

\end{itemize}

\subsection*{Acknowledgements}

We acknowledge funding from NSF ACI-1036211 and NSF AST-1009781. The BlueTides simulation was run on facilities at the National Center for Supercomputing Applications. SMW acknowledge support from the UK Science and Technology Facilities Council through the Sussex Consolidated Grant (ST/L000652/1).

\bibliographystyle{mnras}
\bibliography{wilkins} 

\appendix

\section{Data}

\begin{table*}
\caption{Tabulated values of the stellar to dark matter mass ratio used in Fig. \ref{fig:DM_stellar}.}
\label{tab:DM_stellar}
\begin{tabular}{cccccccc}
\hline
    & \multicolumn{7}{c}{$\log_{10}(M_{\rm DM}/{\rm M_{\odot}})=$} \\
$z$ & $10.5$-$10.75$ & $10.75$-$11.0$ & $11.0$-$11.25$ & $11.25$-$11.5$ & $11.5$-$11.75$ & $11.75$-$12.0$ & $12.0$-$12.25$ \\
\hline
    & \multicolumn{7}{c}{{\bf stellar-to-dark matter mass ratio} - $\log_{10}(M_*/M_{\rm DM})$} \\
\hline
 13.0 & $-2.75 $  & $-2.56 $  & - & - & - & - & -\\
 12.0 & $-2.67 $  & $-2.54 $  & $-2.46 $  & - & - & - & -\\
 11.0 & $-2.66 $  & $-2.52 $  & $-2.38 $  & $-2.28 $  & - & - & -\\
 10.0 & $-2.6 $  & $-2.45 $  & $-2.3 $  & $-2.21 $  & - & - & -\\
 9.0 & $-2.57 $  & $-2.4 $  & $-2.26 $  & $-2.11 $  & $-1.99 $  & $-2.05 $  & -\\
 8.0 & $-2.52 $  & $-2.34 $  & $-2.18 $  & $-2.05 $  & $-1.89 $  & $-1.75 $  & $-1.87 $ \\
\hline
\end{tabular}
\end{table*}

\begin{table*}
\caption{Tabulated values of the galaxy stellar mass function (GSMF) used in Fig. \ref{fig:GSMF}. }
\label{tab:GSMF}
\begin{tabular}{ccccccc}
\hline
    & \multicolumn{6}{c}{$\log_{10}(\phi/{\rm dex^{-1}Mpc^{-3}})$} \\
        \hline
$\log_{10}(M_*/{\rm M_{\odot}})$ & $z=13$ & $z=12$ & $z=11$ & $z=10$ & $z=9$ & $z=8$  \\
\hline
 $ 8.0 $ - $8.2 $  & $ -5.61 $  & $ -4.94 $  & $ -4.31 $  & $ -3.72 $  & $ -3.22 $  & $ -2.76 $ \\
 $ 8.2 $ - $8.4 $  & $ -6.02 $  & $ -5.3 $  & $ -4.64 $  & $ -4.01 $  & $ -3.45 $  & $ -2.97 $ \\
 $ 8.4 $ - $8.6 $  & $ -6.4 $  & $ -5.7 $  & $ -4.97 $  & $ -4.3 $  & $ -3.71 $  & $ -3.19 $ \\
 $ 8.6 $ - $8.8 $  & - & $ -6.14 $  & $ -5.35 $  & $ -4.61 $  & $ -3.98 $  & $ -3.42 $ \\
 $ 8.8 $ - $9.0 $  & - & $ -6.53 $  & $ -5.7 $  & $ -4.96 $  & $ -4.27 $  & $ -3.66 $ \\
 $ 9.0 $ - $9.2 $  & - & - & $ -6.21 $  & $ -5.34 $  & $ -4.58 $  & $ -3.92 $ \\
 $ 9.2 $ - $9.4 $  & - & - & - & $ -5.65 $  & $ -4.93 $  & $ -4.2 $ \\
 $ 9.4 $ - $9.6 $  & - & - & - & $ -6.11 $  & $ -5.28 $  & $ -4.5 $ \\
 $ 9.6 $ - $9.8 $  & - & - & - & - & $ -5.65 $  & $ -4.87 $ \\
 $ 9.8 $ - $10.0 $  & - & - & - & - & $ -6.07 $  & $ -5.17 $ \\
 $ 10.0 $ - $10.2 $  & - & - & - & - & - & $ -5.59 $ \\
 $ 10.2 $ - $10.4 $  & - & - & - & - & - & $ -5.96 $ \\
\hline
\end{tabular}
\end{table*}

\begin{table*}
\caption{Tabulated values of the star formation rate distribution function (SFRDF) used in Fig. \ref{fig:SFRDF}.}
\label{tab:SFRDF}
\begin{tabular}{ccccccc}
\hline
    & \multicolumn{6}{c}{$\log_{10}(\phi/{\rm dex^{-1}Mpc^{-3}})$} \\
    \hline
$\log_{10}({\rm SFR}/{\rm M_{\odot}\,yr^{-1}})$ & $z=13$ & $z=12$ & $z=11$ & $z=10$ & $z=9$ & $z=8$  \\
\hline
 $ -0.6 $ - $ -0.4 $  & $-3.93 $ & $-3.5 $ & $-3.11 $ & $-2.8 $ & $-2.51 $ & $-2.22 $\\
 $ -0.4 $ - $ -0.2 $  & $-4.21 $ & $-3.77 $ & $-3.36 $ & $-3.02 $ & $-2.71 $ & $-2.41 $\\
 $ -0.2 $ - $ 0.0 $  & $-4.54 $ & $-4.07 $ & $-3.62 $ & $-3.26 $ & $-2.93 $ & $-2.61 $\\
 $ 0.0 $ - $ 0.2 $  & $-4.92 $ & $-4.42 $ & $-3.93 $ & $-3.52 $ & $-3.16 $ & $-2.8 $\\
 $ 0.2 $ - $ 0.4 $  & $-5.28 $ & $-4.72 $ & $-4.22 $ & $-3.78 $ & $-3.39 $ & $-3.01 $\\
 $ 0.4 $ - $ 0.6 $  & $-5.67 $ & $-5.13 $ & $-4.54 $ & $-4.06 $ & $-3.64 $ & $-3.22 $\\
 $ 0.6 $ - $ 0.8 $  & $-6.04 $ & $-5.49 $ & $-4.89 $ & $-4.39 $ & $-3.9 $ & $-3.46 $\\
 $ 0.8 $ - $ 1.0 $  & $-6.4 $ & $-5.82 $ & $-5.21 $ & $-4.69 $ & $-4.14 $ & $-3.7 $\\
 $ 1.0 $ - $ 1.2 $  & - & $-6.23 $ & $-5.63 $ & $-5.05 $ & $-4.44 $ & $-3.97 $\\
 $ 1.2 $ - $ 1.4 $  & - & - & $-5.98 $ & $-5.41 $ & $-4.79 $ & $-4.25 $\\
 $ 1.4 $ - $ 1.6 $  & - & - & $-6.32 $ & $-5.77 $ & $-5.12 $ & $-4.57 $\\
 $ 1.6 $ - $ 1.8 $  & - & - & - & $-6.21 $ & $-5.44 $ & $-4.91 $\\
 $ 1.8 $ - $ 2.0 $  & - & - & - & - & $-5.91 $ & $-5.28 $\\
 $ 2.0 $ - $ 2.2 $  & - & - & - & - & $-6.34 $ & $-5.65 $\\
 $ 2.2 $ - $ 2.4 $  & - & - & - & - & - & $-6.12 $\\
 $ 2.4 $ - $ 2.6 $  & - & - & - & - & - & $-6.57 $\\
\hline
\end{tabular}
\end{table*}

\begin{table*}
\caption{Tabulated values of the median specific star formation rate, mass-weighted stellar age, star forming gas metallicity, and stellar metallicity used in Figures \ref{fig:M_sSFR}, {fig:ages} and \ref{fig:metallicities}.}
\label{tab:physical}
\begin{tabular}{cccccccccc}
\hline
    & \multicolumn{9}{c}{$\log_{10}(M_*/{\rm M_{\odot}})=$} \\
$z$ & $8.0$-$8.25$ & $8.25$-$8.5$ & $8.5$-$8.75$ & $8.75$-$9.0$ & $9.0$-$9.25$ & $9.25$-$9.50$ & $9.50$-$9.75$ & $9.75$-$10.0$ & $10.0$-$10.25$ \\
\hline
    & \multicolumn{9}{c}{{\bf median specific star formation rate} - $\log_{10}[({\rm SFR}/M_{*})/{\rm yr^{-1}}]$} \\
\hline
 13.0 & $-7.7 $  & $-7.65 $  & - & - & - & - & - & - & -\\
 12.0 & $-7.76 $  & $-7.77 $  & $-7.74 $  & $-7.74 $  & - & - & - & - & -\\
 11.0 & $-7.81 $  & $-7.81 $  & $-7.79 $  & $-7.76 $  & $-7.74 $  & - & - & - & -\\
 10.0 & $-7.91 $  & $-7.91 $  & $-7.91 $  & $-7.9 $  & $-7.89 $  & $-7.89 $  & $-7.96 $  & - & -\\
 9.0 & $-8.01 $  & $-8.0 $  & $-7.99 $  & $-7.97 $  & $-7.98 $  & $-7.96 $  & $-7.97 $  & $-7.96 $  & -\\
 8.0 & $-8.1 $  & $-8.09 $  & $-8.09 $  & $-8.1 $  & $-8.1 $  & $-8.1 $  & $-8.11 $  & $-8.11 $  & $-8.1 $ \\
\hline
    & \multicolumn{9}{c}{{\bf median mass-weighted age} - $age/{\rm Myr}$} \\
\hline
 13.0 & 33 & 33 & - & - & - & - & - & - & -\\
 12.0 & 39 & 41 & 40 & 40 & - & - & - & - & -\\
 11.0 & 47 & 48 & 48 & 46 & 49 & - & - & - & -\\
 10.0 & 57 & 56 & 56 & 56 & 56 & 56 & 56 & - & -\\
 9.0 & 71 & 71 & 70 & 70 & 70 & 70 & 70 & 71 & -\\
 8.0 & 89 & 88 & 88 & 88 & 87 & 88 & 88 & 89 & 89\\
\hline
    & \multicolumn{9}{c}{{\bf median star forming gas metallicity} - $\log_{10}Z_{\rm SFG}$} \\
\hline
 13.0 & $-2.95 $  & $-2.85 $  & - & - & - & - & - & - & -\\
 12.0 & $-2.99 $  & $-2.86 $  & $-2.75 $  & $-2.62 $  & - & - & - & - & -\\
 11.0 & $-3.0 $  & $-2.89 $  & $-2.79 $  & $-2.65 $  & $-2.56 $  & - & - & - & -\\
 10.0 & $-3.02 $  & $-2.91 $  & $-2.81 $  & $-2.71 $  & $-2.58 $  & $-2.5 $  & $-2.38 $  & - & -\\
 9.0 & $-3.02 $  & $-2.92 $  & $-2.82 $  & $-2.71 $  & $-2.61 $  & $-2.5 $  & $-2.39 $  & $-2.26 $  & -\\
 8.0 & $-3.03 $  & $-2.92 $  & $-2.82 $  & $-2.72 $  & $-2.61 $  & $-2.5 $  & $-2.4 $  & $-2.27 $  & $-2.17 $ \\
\hline
    & \multicolumn{9}{c}{{\bf median stellar metallicity} - $\log_{10}Z_{*}$} \\
\hline
 13.0 & $-3.11 $  & $-3.01 $  & - & - & - & - & - & - & -\\
 12.0 & $-3.15 $  & $-3.04 $  & $-2.92 $  & $-2.82 $  & - & - & - & - & -\\
 11.0 & $-3.17 $  & $-3.06 $  & $-2.95 $  & $-2.83 $  & $-2.75 $  & - & - & - & -\\
 10.0 & $-3.17 $  & $-3.07 $  & $-2.98 $  & $-2.87 $  & $-2.75 $  & $-2.66 $  & $-2.56 $  & - & -\\
 9.0 & $-3.18 $  & $-3.08 $  & $-2.98 $  & $-2.88 $  & $-2.78 $  & $-2.67 $  & $-2.57 $  & $-2.45 $  & -\\
 8.0 & $-3.19 $  & $-3.09 $  & $-2.99 $  & $-2.89 $  & $-2.79 $  & $-2.68 $  & $-2.58 $  & $-2.47 $  & $-2.38 $ \\
\hline
\end{tabular}
\end{table*}

\begin{table*}
\caption{Tabulated values of the median far-UV photon escape fraction in stellar mass bins used in Fig. \ref{fig:L_fesc}.}
\label{tab:L_fesc}
\begin{tabular}{cccccccccc}
\hline
    & \multicolumn{9}{c}{$\log_{10}(M_*/{\rm M_{\odot}})=$} \\
$z$ & $8.0$-$8.25$ & $8.25$-$8.5$ & $8.5$-$8.75$ & $8.75$-$9.0$ & $9.0$-$9.25$ & $9.25$-$9.50$ & $9.50$-$9.75$ & $9.75$-$10.0$ & $10.0$-$10.25$ \\
\hline
    & \multicolumn{9}{c}{{\bf median far-UV photon escape fraction}} \\
\hline
 13.0 & 0.87 & 0.77 & - & - & - & - & - & - & -\\
 12.0 & 0.91 & 0.78 & 0.6 & 0.43 & - & - & - & - & -\\
 11.0 & 0.9 & 0.79 & 0.62 & 0.39 & 0.32 & - & - & - & -\\
 10.0 & 0.91 & 0.8 & 0.67 & 0.5 & 0.33 & 0.26 & 0.18 & - & -\\
 9.0 & 0.9 & 0.8 & 0.65 & 0.49 & 0.35 & 0.24 & 0.18 & 0.14 & -\\
 8.0 & 0.89 & 0.8 & 0.67 & 0.52 & 0.38 & 0.27 & 0.2 & 0.15 & 0.12\\
\hline
\end{tabular}
\end{table*}

\begin{table*}
\caption{Tabulated values of the median intrinsic and observed far-UV mass-to-light ratio used in Fig. \ref{fig:MTOL}.}
\label{tab:MTOL}
\begin{tabular}{cccccccccc}
\hline
    & \multicolumn{9}{c}{$\log_{10}(M_*/{\rm M_{\odot}})=$} \\
$z$ & $8.0$-$8.25$ & $8.25$-$8.5$ & $8.5$-$8.75$ & $8.75$-$9.0$ & $9.0$-$9.25$ & $9.25$-$9.50$ & $9.50$-$9.75$ & $9.75$-$10.0$ & $10.0$-$10.25$ \\
\hline
    & \multicolumn{9}{c}{{\bf median intrinsic mass-to-light ratio} -  $\log_{10}[(M_{*}/{\rm M_{\odot}})/(L_{\nu, {\rm fuv}}/{\rm erg s^{-1} Hz^{-1}})]$} \\
\hline
 13.0 & $ -20.42 $  & $ -20.44 $  & - & - & - & - & - & - & -\\
 12.0 & $ -20.38 $  & $ -20.34 $  & $ -20.37 $  & $ -20.37 $  & - & - & - & - & -\\
 11.0 & $ -20.3 $  & $ -20.3 $  & $ -20.31 $  & $ -20.33 $  & $ -20.37 $  & - & - & - & -\\
 10.0 & $ -20.25 $  & $ -20.25 $  & $ -20.25 $  & $ -20.24 $  & $ -20.23 $  & $ -20.23 $  & $ -20.21 $  & - & -\\
 9.0 & $ -20.15 $  & $ -20.16 $  & $ -20.17 $  & $ -20.17 $  & $ -20.17 $  & $ -20.17 $  & $ -20.18 $  & $ -20.17 $  & -\\
 8.0 & $ -20.08 $  & $ -20.09 $  & $ -20.09 $  & $ -20.09 $  & $ -20.08 $  & $ -20.08 $  & $ -20.07 $  & $ -20.06 $  & $ -20.08 $ \\
\hline
    & \multicolumn{9}{c}{{\bf median observed mass-to-light ratio} -  $\log_{10}[(M_{*}/{\rm M_{\odot}})/(L_{\nu, {\rm fuv}}/{\rm erg s^{-1} Hz^{-1}})]$} \\
\hline
 13.0 & $-20.37 $  & $-20.33 $  & - & - & - & - & - & - & -\\
 12.0 & $-20.32 $  & $-20.23 $  & $-20.15 $  & $-20.0 $  & - & - & - & - & -\\
 11.0 & $-20.25 $  & $-20.19 $  & $-20.1 $  & $-19.94 $  & $-19.81 $  & - & - & - & -\\
 10.0 & $-20.2 $  & $-20.15 $  & $-20.07 $  & $-19.93 $  & $-19.77 $  & $-19.62 $  & $-19.5 $  & - & -\\
 9.0 & $-20.1 $  & $-20.05 $  & $-19.97 $  & $-19.86 $  & $-19.72 $  & $-19.56 $  & $-19.44 $  & $-19.32 $  & -\\
 8.0 & $-20.03 $  & $-19.98 $  & $-19.91 $  & $-19.8 $  & $-19.67 $  & $-19.51 $  & $-19.38 $  & $-19.25 $  & $-19.21 $ \\
\hline
\end{tabular}
\end{table*}

\begin{table*}
\caption{Tabulated values of the intrinsic and observed (dust-attenuated) far-UV luminosity functions used in Figure \ref{fig:UVLF}.}
\label{tab:UVLF}
\begin{tabular}{ccccccc}
\hline
    & \multicolumn{6}{c}{$\log_{10}(\phi/{\rm mag^{-1}Mpc^{-3}})$} \\
        \hline
$M_{\rm fuv}$ & $z=13$ & $z=12$ & $z=11$ & $z=10$ & $z=9$ & $z=8$  \\
\hline
 \multicolumn{7}{c}{{\bf intrinsic far-UV luminosity function}} \\
\hline
 $ -25.0$ - $-24.5 $  & - & - & - & - & - & $-6.71$\\
 $ -24.5$ - $-24.0 $  & - & - & - & - & - & $-6.26$\\
 $ -24.0$ - $-23.5 $  & - & - & - & - & $-6.54$ & $-5.89$\\
 $ -23.5$ - $-23.0 $  & - & - & - & - & $-6.19$ & $-5.46$\\
 $ -23.0$ - $-22.5 $  & - & - & - & $-6.37$ & $-5.71$ & $-5.13$\\
 $ -22.5$ - $-22.0 $  & - & - & $-6.63$ & $-6.04$ & $-5.36$ & $-4.77$\\
 $ -22.0$ - $-21.5 $  & - & - & $-6.35$ & $-5.66$ & $-5.05$ & $-4.48$\\
 $ -21.5$ - $-21.0 $  & - & $-6.63$ & $-5.92$ & $-5.29$ & $-4.71$ & $-4.21$\\
 $ -21.0$ - $-20.5 $  & $-6.82$ & $-6.15$ & $-5.54$ & $-4.94$ & $-4.44$ & $-3.96$\\
 $ -20.5$ - $-20.0 $  & $-6.43$ & $-5.77$ & $-5.22$ & $-4.64$ & $-4.17$ & $-3.71$\\
 $ -20.0$ - $-19.5 $  & $-5.96$ & $-5.37$ & $-4.87$ & $-4.33$ & $-3.91$ & $-3.49$\\
 $ -19.5$ - $-19.0 $  & $-5.6$ & $-5.04$ & $-4.56$ & $-4.04$ & $-3.66$ & $-3.26$\\
 $ -19.0$ - $-18.5 $  & $-5.18$ & $-4.66$ & $-4.23$ & $-3.77$ & $-3.42$ & $-3.05$\\
 $ -18.5$ - $-18.0 $  & $-4.84$ & $-4.33$ & $-3.92$ & $-3.5$ & $-3.19$ & $-2.84$\\
 $ -18.0$ - $-17.5 $  & $-4.46$ & $-3.99$ & $-3.63$ & $-3.25$ & $-2.98$ & $-2.64$\\
 $ -17.5$ - $-17.0 $  & $-4.12$ & $-3.69$ & $-3.37$ & $-3.02$ & $-2.77$ & $-2.44$\\
\hline
 \multicolumn{7}{c}{{\bf observed (dust-corrected) far-UV luminosity function}} \\
\hline
 $ -23.0$ - $-22.5 $  & - & - & - & - & - & $-6.89$\\
 $ -22.5$ - $-22.0 $  & - & - & - & - & $-6.61$ & $-6.07$\\
 $ -22.0$ - $-21.5 $  & - & - & - & $-6.59$ & $-6.06$ & $-5.35$\\
 $ -21.5$ - $-21.0 $  & - & - & $-6.52$ & $-5.91$ & $-5.37$ & $-4.8$\\
 $ -21.0$ - $-20.5 $  & - & $-6.49$ & $-5.91$ & $-5.28$ & $-4.76$ & $-4.27$\\
 $ -20.5$ - $-20.0 $  & $-6.59$ & $-5.9$ & $-5.33$ & $-4.73$ & $-4.29$ & $-3.85$\\
 $ -20.0$ - $-19.5 $  & $-5.98$ & $-5.39$ & $-4.88$ & $-4.35$ & $-3.93$ & $-3.5$\\
 $ -19.5$ - $-19.0 $  & $-5.6$ & $-5.01$ & $-4.55$ & $-4.04$ & $-3.65$ & $-3.25$\\
 $ -19.0$ - $-18.5 $  & $-5.17$ & $-4.66$ & $-4.23$ & $-3.76$ & $-3.41$ & $-3.03$\\
 $ -18.5$ - $-18.0 $  & $-4.84$ & $-4.32$ & $-3.92$ & $-3.49$ & $-3.18$ & $-2.82$\\
 $ -18.0$ - $-17.5 $  & $-4.45$ & $-3.99$ & $-3.63$ & $-3.25$ & $-2.97$ & $-2.63$\\
 $ -17.5$ - $-17.0 $  & $-4.12$ & $-3.69$ & $-3.37$ & $-3.02$ & $-2.77$ & $-2.44$\\
\hline
\end{tabular}
\end{table*}

\begin{table*}
\caption{Median SMBH masses in stellar mass bins used in Fig. \ref{fig:M_MSMBH}.}
\label{tab:SMBH}
\begin{tabular}{cccccccccc}
\hline
    & \multicolumn{9}{c}{$\log_{10}(M_*/{\rm M_{\odot}})=$} \\
$z$ & $8.0$-$8.25$ & $8.25$-$8.5$ & $8.5$-$8.75$ & $8.75$-$9.0$ & $9.0$-$9.25$ & $9.25$-$9.50$ & $9.50$-$9.75$ & $9.75$-$10.0$ & $10.0$-$10.25$ \\
\hline
    & \multicolumn{9}{c}{{\bf median SMBH mass} - $\log_{10}(M_{\rm SMBH}/M_{\odot})$} \\
\hline
 13.0 & - & - & - & - & - & - & - & - & -\\
 12.0 & - & - & - & 5.93 & - & - & - & - & -\\
 11.0 & - & - & - & 5.95 & 6.17 & - & - & - & -\\
 10.0 & - & - & - & 5.96 & 6.19 & 6.49 & 6.69 & - & -\\
 9.0 & - & - & - & 5.97 & 6.18 & 6.44 & 6.68 & 6.99 & -\\
 8.0 & - & - & 5.86 & 5.98 & 6.17 & 6.4 & 6.65 & 6.94 & 7.12\\
 \hline
\end{tabular}
\end{table*}

\end{document}